\documentclass[journal,10pt]{IEEEtran}
\usepackage[figurename=Fig.]{caption}
\usepackage{balance}
\usepackage{graphicx} 
\usepackage{tikz}
\usetikzlibrary{decorations.pathreplacing, calc}
\usepackage{algorithm}
\usepackage{algpseudocode}
\usepackage{titlesec}
\usepackage{float}
\usepackage{filecontents} 
\usetikzlibrary{calc, positioning, shapes.geometric, decorations.pathreplacing}
\usepackage{array}
\usepackage{verbatim}
\usetikzlibrary{calc, shapes, arrows, positioning}
\usepackage{authblk}
\usepackage{blindtext}
\usepackage{filecontents} 
\usepackage{algorithmicx} 
\usepackage{ifpdf}
\usepackage{graphicx}
\usepackage{tabulary}
\usepackage{amsmath,amsthm,amssymb}
\usepackage{amsmath}

\usepackage{kantlipsum}
\allowdisplaybreaks
\usepackage{cleveref}
\usepackage{lipsum}%
\usepackage{optidef}
\usepackage[english]{babel} 
\theoremstyle{plain}
\newtheorem{theorem}{Theorem}

\crefname{theorem}{Theorem}{theorem}
\crefname{lemma}{Lemma}{Lemmas}

\usepackage{cite}
\usepackage{dsfont}
\usepackage{dblfloatfix}
\usepackage[justification=centering]{caption}
\usepackage{xcolor,cite,etoolbox}
\makeatletter 
\pretocmd\@bibitem{\color{black}\csname keycolor#1\endcsname}{}{\fail}
\newcommand\citecolor[1]{\@namedef{keycolor#1}{\color{black}}}
\usepackage{algorithm}
\usepackage{algorithmicx, algpseudocode}
\usepackage{algpseudocode}
\algrenewcommand{\algorithmicdo}{}    
\algrenewcommand{\algorithmicthen}{}  
\usepackage{etoolbox}
\usepackage{subcaption}
\usepackage{balance}
\usepackage{empheq,etoolbox}
\usepackage[utf8]{inputenc}
\usepackage{multirow}
\usepackage{float} 
\usepackage{amssymb}
\usepackage{pifont}
\floatstyle{plaintop}
\restylefloat{table}
\patchcmd{\subequations}
\usepackage{lipsum} 

\usepackage{lipsum} 

\tikzset{brace/.style={decorate, decoration={brace}},
 brace mirrored/.style={decorate, decoration={brace,mirror}},
}

\newcounter{brace}
\newcounter{arrow}
\citecolor{mu2021simultaneously}
\citecolor{10032506}
\citecolor{10535263}
\citecolor{10480441}
\citecolor{10844052}
\citecolor{10466748}
\citecolor{10502274}
\citecolor{10028982}
\citecolor{6868214}
\citecolor{6786060}

\begin{document}

 \captionsetup[figure]{name={Fig.},labelsep=period}

 \title{Efficient STAR-RIS Mode for Energy Minimization in WPT-FL Networks with NOMA}

\author{
MohammadHossein Alishahi, Ming Zeng, Paul Fortier, Omer Waqar, Muhammad Hanif, Dinh Thai Hoang, Diep N. Nguyen, and Quoc-Viet Pham
\thanks{
This work was supported in part by NSERC of Canada under Grants RGPIN-2021-02636 and CRC-2022-00115, and by FRQNT through its early research program under Grant 341270. (Corresponding author: Ming Zeng.) 

M. H. Alishahi, M. Zeng, and P. Fortier are with the Department of Electrical and Computer Engineering, Laval University, Quebec City, QC, G1V 0A6, CA. email:  mohammadhossein.alishahi.1@ulaval.ca;  \{ming.zeng, paul.fortier\}@gel.ulaval.ca.}

\thanks{Omer Waqar is with the School of Computing, University of the Fraser Valley, Abbotsford, BC, V2S 7M8, CA. email: Omer.Waqar@ufv.ca.}

\thanks{Muhammad Hanif is with the Department of Engineering, Thompson Rivers University, Kamloops, BC, V2C 0C8, Canada. email: mhanif@tru.ca.}

\thanks{Dinh~Thai~Hoang and Diep~N.~Nguyen are with the School of Electrical and Data Engineering, University of Technology Sydney, Sydney, NSW 2007, Australia e-mail: \{hoang.dinh, diep.nguyen\}@uts.edu.au.}

\thanks{Quoc-Viet Pham is with the School of Computer Science and Statistics, Trinity College Dublin, Dublin 2, D02PN40, Ireland. email: viet.pham@tcd.ie.}}

\maketitle

\begin{abstract}
With the massive deployment of Internet of Things (IoT) devices in sixth-generation networks, several critical challenges have emerged, such as large communication overhead, coverage limitations, and limited battery lifespan due to high energy consumption. Federated learning (FL), wireless power transfer (WPT), multi-antenna access point (AP), and reconfigurable intelligent surfaces (RIS) can mitigate these challenges by reducing the need for large data transmissions, enabling sustainable energy harvesting, and optimizing the propagation environment. Compared to conventional RIS, simultaneously transmitting and reflecting (STAR)-RIS not only extends coverage from half-space to full-space but also improves energy saving through appropriate mode selection. Motivated by the need for sustainable, low-latency, and energy-efficient communication in large-scale IoT networks, this paper investigates the efficient STAR-RIS mode in the uplink and downlink phases of a WPT-FL multi-antenna AP network with non-orthogonal multiple access to minimize energy consumption,  a joint optimization that remains largely unexplored in existing works on RIS or STAR-RIS. We formulate a non-convex energy minimization problem for different STAR-RIS modes, i.e., energy splitting (ES) and time switching (TS), in both uplink and downlink transmission phases, where STAR-RIS phase shift vectors, beamforming matrices, time and power for harvesting, uplink transmission, and downlink transmission, local processing time, and computation frequency for each user are jointly optimized. To tackle the non-convexity, the problem is decoupled into two subproblems: the first subproblem optimizes STAR-RIS phase shift vectors and beamforming matrices across all WPT-FL phases using block coordinate descent over either semi-definite programming or Rayleigh quotient problems, while the second one allocates time, power, and computation frequency via the one-dimensional search algorithms or the bisection algorithm. Simulation results demonstrate that TS STAR-RIS in both uplink and downlink transmissions achieves the lowest energy consumption, outperforming ES and conventional RIS schemes due to its flexible phase shift adaptation and lower interference levels.
\end{abstract}
\vspace{-0.1cm}

\section{Introduction}
\IEEEPARstart{M}{assive} deployment of Internet of Things (IoT) devices is 
poised to be a cornerstone of beyond fifth-generation (B5G) and sixth-generation (6G) wireless networks, enabling intelligent services across diverse domains, such as smart cities, autonomous vehicles, and industrial automation \cite{10597395}. However, large-scale IoT networks face several challenges, including large communication overhead, coverage limitations, and limited battery lifespan due to high energy consumption \cite{10387440}.

To address limited battery lifespan, wireless power transfer (WPT) has emerged as a key enabler, allowing battery-constrained IoT devices to harvest energy wirelessly and prolong their operational lifespan without frequent battery replacements \cite{agbinya2022wireless}. In parallel, technologies such as non-orthogonal multiple access (NOMA) and multi-antenna access points (APs) have been leveraged to reduce energy consumption and enhance scalability and spectral efficiency \cite{8319526, apiyo2024survey}. Specifically, NOMA improves system capacity by enabling simultaneous communication for multiple users, while multi-antenna APs use efficient beamforming to combat channel attenuation and reduce energy consumption \cite{8319526, apiyo2024survey}. Moreover, federated learning (FL) has gained attention for mitigating communication overhead, reducing energy consumption, and addressing privacy concerns by decentralizing model training, allowing IoT devices to process and update models locally \cite{10420449}.

Despite these advancements, large-scale IoT deployments often encounter coverage gaps and unreliable links due to obstacles and unfavorable channel conditions. Reconfigurable intelligent surfaces (RIS) have been proposed to manipulate the wireless propagation environment dynamically, thus enhancing signal strength and coverage by adjusting the phase shifts of passive reflective elements and addressing the challenge of coverage limitation in IoT networks\cite{10496996, 9729826}. However, conventional RIS is inherently limited to reflecting signals in one direction, restricting coverage flexibility \cite{9729826, 10133841}. To overcome this, simultaneous transmitting and reflecting (STAR)-RIS has been introduced, offering full coverage by simultaneously transmitting and reflecting signals \cite{10133841}. STAR-RIS operates in three modes, energy splitting (ES), mode switching (MS), and time switching (TS), allowing flexible control over signal propagation in various scenarios \cite{mu2021simultaneously}.

Motivated by these advancements, this work integrates WPT, multi-antenna APs, and STAR-RIS into an FL framework with NOMA to jointly address energy consumption, large communication overhead, and coverage challenges in large-scale IoT networks. Furthermore, identifying the efficient STAR-RIS operation mode in each phase of the WPT-FL network can further reduce energy consumption. Therefore, we investigate the best STAR-RIS operation mode in each phase of the WPT-FL network with NOMA to achieve minimal energy consumption.
{\color{blue}

\begin{table*}[htbp]
    \centering
  
    \begin{tabular}{|>{\centering\arraybackslash}m{1.6cm}|
                    >{\centering\arraybackslash}m{3.1cm}|
                    >{\centering\arraybackslash}m{0.9cm}|
                    >{\centering\arraybackslash}m{0.3cm}|
                    >{\centering\arraybackslash}m{0.4cm}|
                    >{\centering\arraybackslash}m{0.3cm}|
                    >{\centering\arraybackslash}m{0.6cm}|
                    >{\centering\arraybackslash}m{0.9cm}|
                    >{\centering\arraybackslash}m{0.9cm}|
                    >{\centering\arraybackslash}m{0.6cm}|
                    >{\centering\arraybackslash}m{1.1cm}|
                    >{\centering\arraybackslash}m{1.2cm}|
                    >{\centering\arraybackslash}m{1cm}|}
    \hline
    Ref. & Performance Metric & \multicolumn{2}{c|}{\textbf{STAR-RIS}} & RIS & FL & WPT & SIC NOMA & Multi-antenna & \multicolumn{4}{c|}{\textbf{Optimized Resources}} \\ 
    \cline{3-4} \cline{10-13}
    & & ES/MS & TS & & & & & & WPT & Uplink & Downlink & Training  \\ \hline
    \cite{9844152} & Energy Efficiency & - & - & - & \checkmark & - & - & - & - & - & - & \checkmark \\ \hline
    \cite{10123399} & Energy Efficiency & - & - & - & \checkmark & - & \checkmark & - & - & \checkmark & - & - \\ \hline 
    \cite{10285339} & Latency & - & - & - & \checkmark & \checkmark & - & \checkmark & \checkmark & \checkmark & - & \checkmark \\ \hline
    \cite{9748877} & Energy Consumption & - & - & - & \checkmark & \checkmark & \checkmark & - & \checkmark & \checkmark & - & - \\ \hline  
    \cite{10044972} & Energy Consumption & - & - & - & \checkmark & \checkmark & \checkmark & - & \checkmark & \checkmark & \checkmark & - \\ \hline  
    \cite{10293758} & Latency & - & - & - & \checkmark & \checkmark & \checkmark & - & \checkmark & \checkmark & \checkmark & - \\ \hline 
    \cite{9857928} & Consumed Energy & - & - & \checkmark & \checkmark & \checkmark & - & - & \checkmark & \checkmark & - & \checkmark  \\ \hline 
    \cite{joshi2024federated} & Energy Efficiency & - & - & \checkmark & \checkmark & \checkmark & - & - & \checkmark & \checkmark & - & - \\ \hline 
    \cite{9798759} & Sum Rate & - & - & \checkmark & \checkmark & - & \checkmark & - & - & \checkmark & - & \checkmark \\ \hline
    \cite{10400811} & Energy Consumption & - & - & \checkmark & \checkmark & \checkmark & \checkmark & - & \checkmark & \checkmark & \checkmark & - \\ \hline 
    \cite{10680285} & Energy Consumption & - & - & \checkmark & \checkmark & - & - & - & - & \checkmark & \checkmark & \checkmark \\ \hline 
    \cite{10572362} & Spectral Efficiency & \checkmark & - & - & \checkmark & \checkmark & - & \checkmark & \checkmark & \checkmark & - & \checkmark \\ \hline
    \cite{10177732} & Spectral Efficiency & \checkmark & \checkmark & - & - & - & - & \checkmark & - & \checkmark & - & - \\ \hline
    \cite{10032506} & Computational bits & \checkmark & \checkmark & - & - & \checkmark & - & - & \checkmark & \checkmark & - & - \\ \hline
    \cite{10720523} & Computational bits & \checkmark & \checkmark & - & - & \checkmark & \checkmark & - & \checkmark & \checkmark & - & - \\ \hline
    \cite{9815289} & Sum Rate & \checkmark & - & - & \checkmark & - & \checkmark & \checkmark & - & \checkmark & - & - \\ \hline
    \cite{10032501} & Latency & \checkmark & - & - & \checkmark & \checkmark & - & \checkmark & \checkmark & \checkmark & \checkmark & - \\ \hline
    \cite{10049110} & Sum Rate & \checkmark & - & - & \checkmark & - & \checkmark & \checkmark & - & - & \checkmark & \checkmark \\ \hline
    This Study & Energy Consumption & \checkmark & \checkmark & - & \checkmark & \checkmark & \checkmark & \checkmark & \checkmark & \checkmark & \checkmark & - \\ \hline
    \end{tabular}
    \caption{Comparison of related references}
    \label{tab:1}
\end{table*}}

\subsection{Related Works}

The convergence of WPT and NOMA within FL-based networks has been widely recognized as a transformative approach to overcome the challenges of limited battery lifespan and high energy consumption in large-scale IoT networks, with prior studies demonstrating significant performance improvements in energy-constrained IoT systems \cite{9844152, 10123399, 10285339, 9748877, 10044972, 10293758}.

To overcome signal degradation and coverage limitations in urban areas for B5G and 6G networks, recent research efforts have focused on deploying RIS in such systems \cite{9857928, joshi2024federated, 9798759, 10400811, 10680285}. For instance, \cite{9857928} minimized total transmit power in WPT-FL networks by optimizing transmission time, power control, and RIS phase shifts, with numerical results verifying fast convergence and the benefits of RIS. In parallel, \cite{joshi2024federated} maximized energy efficiency in RIS-aided WPT networks using multi-agent federated reinforcement learning and optimizing energy harvesting and user trajectory, leading to superior energy efficiency compared to systems without RIS. Aiming to fully utilize the benefits brought by multiple AP, NOMA, and RIS in FL-based networks. Zhong \textit{et al.} \cite{9798759} maximized the sum rate for a multi-antenna AP FL network with NOMA, demonstrating the effectiveness of employing multi-antenna AP, RIS, and NOMA in such a system. In terms of energy consumption in FL-based networks, our previous work \cite{10400811} minimized the total energy consumption for the WPT-FL network with NOMA, focusing on the communication part by optimizing resources across the harvesting, uplink, and downlink transmission phases, where the efficiency of employing RIS and the superiority of NOMA compared to frequency division multiple access (FDMA) is revealed. Considering the training part, \cite{10680285} minimized energy consumption in RIS-aided FL networks by optimizing all resources in FL phases, achieving significant energy savings and near-interference-free learning through the derived minimum global communication and local iteration rounds for efficient FL aggregation. 

More recently, researchers have actively developed STAR-RIS as a breakthrough technology in this context, adding full coverage on top of regular RIS capabilities \cite{10572362, 10177732, 10720523, 10032506, 9815289,10032501, 10049110}. For instance, Hashempour \textit{et al.} \cite{10572362} introduced a secure STAR-RIS aided system for multi-antenna AP downlink networks with rate-splitting multiple access (RSMA), demonstrating that STAR-RIS enhances both sum secrecy rate and overall spectral efficiency compared to conventional RIS. The authors of \cite{10177732} maximized spectral and energy efficiency in STAR-RIS aided interference-limited systems with finite block length by optimizing the reflecting coefficients, demonstrating that STAR-RIS outperforms regular RIS in coverage, with the ES scheme achieving better results than both MS and TS schemes. Qin \textit{et al.} \cite{10032506} demonstrated that the TS mode in STAR-RIS-aided uplink WPT mobile edge computing (MEC) networks with NOMA achieves higher total computational bits than the ES, MS protocols, and traditional RIS. In parallel, our previous study \cite{10720523} identified the best uplink scheme for ES STAR-RIS aided WPT-MEC networks, where time division multiple access (TDMA) computes more bits than other schemes such as hybrid TDMA-NOMA and NOMA, owing to its ability to adapt STAR-RIS phase shift vectors flexibly. 
In FL-based systems, \cite{9815289} integrated STAR-RIS and NOMA into over-the-air FL networks, improving interference mitigation, coverage, convergence, and spectrum efficiency compared to the baselines. Our earlier work \cite{10032501} minimized latency in a multi-antenna ES STAR-RIS aided WPT-FL network, showing that integrating multi-antenna AP, STAR-RIS, and FL effectively reduces latency and expands coverage in IoT networks. The authors of \cite{10049110} proposed a deep reinforcement learning agent with an access-free FL framework in a STAR-RIS-aided downlink NOMA system, achieving enhanced sum rate with ES STAR-RIS over MS STAR-RIS and reduced training overhead compared to traditional methods. 

 However, bringing these theoretical advances into practice faces several challenges. These include hardware constraints, such as the need for affordable high-resolution phase shifters. Another challenge is the complexity of jointly optimizing transmission and reflection coefficients. Maintaining aligned STAR-RIS phase shifts in dynamic environments, especially with mobile IoT deployments, also introduces latency and synchronization issues. Additionally, frequent FL updates can create communication overhead. Finally, imperfect CSI can negatively impact both energy efficiency and model convergence. Recent works have investigated hardware impairment challenges and the impact of imperfect CSI in practical RIS and holographic meta-surface systems. For instance, \cite{10535263} proposed a multi-layer meta-surface architecture that improves spectral and energy efficiency by optimizing layer coefficients under hardware impairments. \cite{10480441} developed an energy-efficient beamforming framework for reconfigurable holographic surfaces, accounting for hardware impairments. Li \textit{et al.} \cite{10844052} designed a hybrid beamforming scheme for low-earth orbit satellite networks that mitigates inter-satellite interference and considers mutual coupling. The authors of \cite{10502274} examined near-field spectral efficiency under phase shift errors and hardware impairments, showing performance gains by increasing base stations. Finally, \cite{10466748} analyzed the effect of imperfect channel state information and hardware impairments on spectral efficiency in intelligent omni-surface systems, showing degradation at high transmit power partly offset by adding more AP antennas. While these studies have advanced our understanding of hardware limitations in RIS and related systems, fully addressing these impairments, particularly in dynamic and large-scale STAR-RIS-assisted FL networks, is beyond the scope of this paper, and we have left it as an interesting direction for future work.

As highlighted in Table~\ref{tab:1}, this work differs from existing studies by jointly integrating STAR-RIS, WPT, FL, multi-antenna APs, and NOMA into a single framework to reduce energy consumption in large-scale IoT networks. Rather than optimizing only a limited set of resources, we address a comprehensive joint optimization problem covering the phase shift vectors, beamforming matrices, power allocation, transmission time across all phases, and computing frequency for each user. Moreover, unlike earlier works examining STAR-RIS in isolation, we investigate the best STAR-RIS operation mode across uplink and downlink transmission phases within a WPT-FL network with NOMA, an aspect not explored before.

\subsection{Motivations and Contributions}
The integration of IoT devices into B5G and 6G networks is challenged by limited battery lifespan due to high energy consumption, as well as large communication overhead and coverage limitations. As highlighted earlier, technologies such as FL, WPT, multi-antenna AP, NOMA, and STAR-RIS can significantly enhance IoT network performance by reducing communication overhead, expanding coverage, and, most critically, reducing energy consumption in such networks. Additionally, identifying the efficient operation mode of STAR-RIS could unlock even greater system performance. Despite advances in these areas, the best STAR-RIS operation mode that minimizes energy consumption in WPT-FL multi-antenna AP networks with NOMA remains largely under-explored. 

This paper investigates the efficient STAR-RIS operation mode for the uplink and downlink transmission of WPT-FL multi-antenna AP networks with NOMA, aiming to address the critical challenges of energy consumption and coverage in IoT systems. The key contributions of this paper are:

\begin{itemize}
    \item For different STAR-RIS operation modes in uplink and downlink transmission phases of the WPT-FL multi-antenna AP system with NOMA, we formulate a non-convex energy minimization problem to jointly optimize the STAR-RIS phase shift vectors, beamforming matrices, power allocation, time across all phases, local processing time, and computing frequency for each user.

    \item The transmission and reflection phase shift vectors of STAR-RIS and the beamforming matrices for energy harvesting and downlink transmission phases are optimized using block coordinate descent (BCD) over formulated semi-definite programming (SDP) problems. Meanwhile, the uplink transmission optimization of the phase shift vectors of STAR-RIS and the beamforming matrix is addressed through BCD applied to the sum of Rayleigh quotients (RQ) and SDP optimization problems. Then, the one-dimensional search algorithm (ODSA) optimizes the harvesting time, local processing time, uplink transmission time, power allocation, and computation frequency for each user across all scenarios. For the obtained resources, another ODSA optimizes the downlink transmission time and power in scenarios with ES STAR-RIS in downlink transmission, while a combination of the bisection algorithm (BA) and ODSA is employed to optimize the downlink resources in scenarios with TS STAR-RIS in downlink transmission.

\item Simulation results indicate that operating STAR-RIS in TS mode for both uplink and downlink transmission enhances energy saving, as it experiences less interference and offers greater flexibility in adapting STAR-RIS phase shift vectors for users compared to ES STAR-RIS. More specifically, the STAR-RIS protocol in uplink transmission plays a crucial role in energy consumption, as the energy required for downlink transmission is negligible compared to the energy consumed for local processing and uplink transmission. Furthermore, STAR-RIS outperforms conventional reflecting-only/transmitting-only RIS, which behaves similarly to MS STAR-RIS, confirming that MS is the least effective protocol for STAR-RIS.
            
\end{itemize}

\begin{table}[htbp]
\centering
    \caption{Notation Table}
    \label{tab:I}
    \begin{tabular}{>{\centering\arraybackslash}p{1.5cm}>{\centering\arraybackslash}p{6.2cm}}
        \hline
        \textbf{Symbol} & \textbf{Meaning} \\
        \hline
        $(.)^\intercal$ & Transposition of a matrix or vector \\
        $(.)^H$ & Hermitian of a matrix or vector \\
        $|.|$ & The cardinality of a set/scalar \\
        $||.||_F$ & Frobenius norm of a matrix \\
        $\mathcal{O}$ & Computational complexity \\
        $\textbf{A} \succeq 0$ & Semi-definite positive matrix \\
        $\mathbb C^{a \times b}$ & Set complex matrix of size $a \times b$ \\
        $\text{diag}(\textbf{a})$ & Diagonal matrix from vector $\textbf{a}$ \\
        \hline
    \end{tabular}
     \vspace{1 em}
    
    \textit{Notations}: Bold lowercase letters denote column vectors, and bold uppercase letters represent matrices. Other notations are listed in Table \ref{tab:I}.
\end{table}

\section{{System Model}}
\begin{figure}[htbp]
\centering
  \includegraphics[keepaspectratio, width=0.52\textwidth]{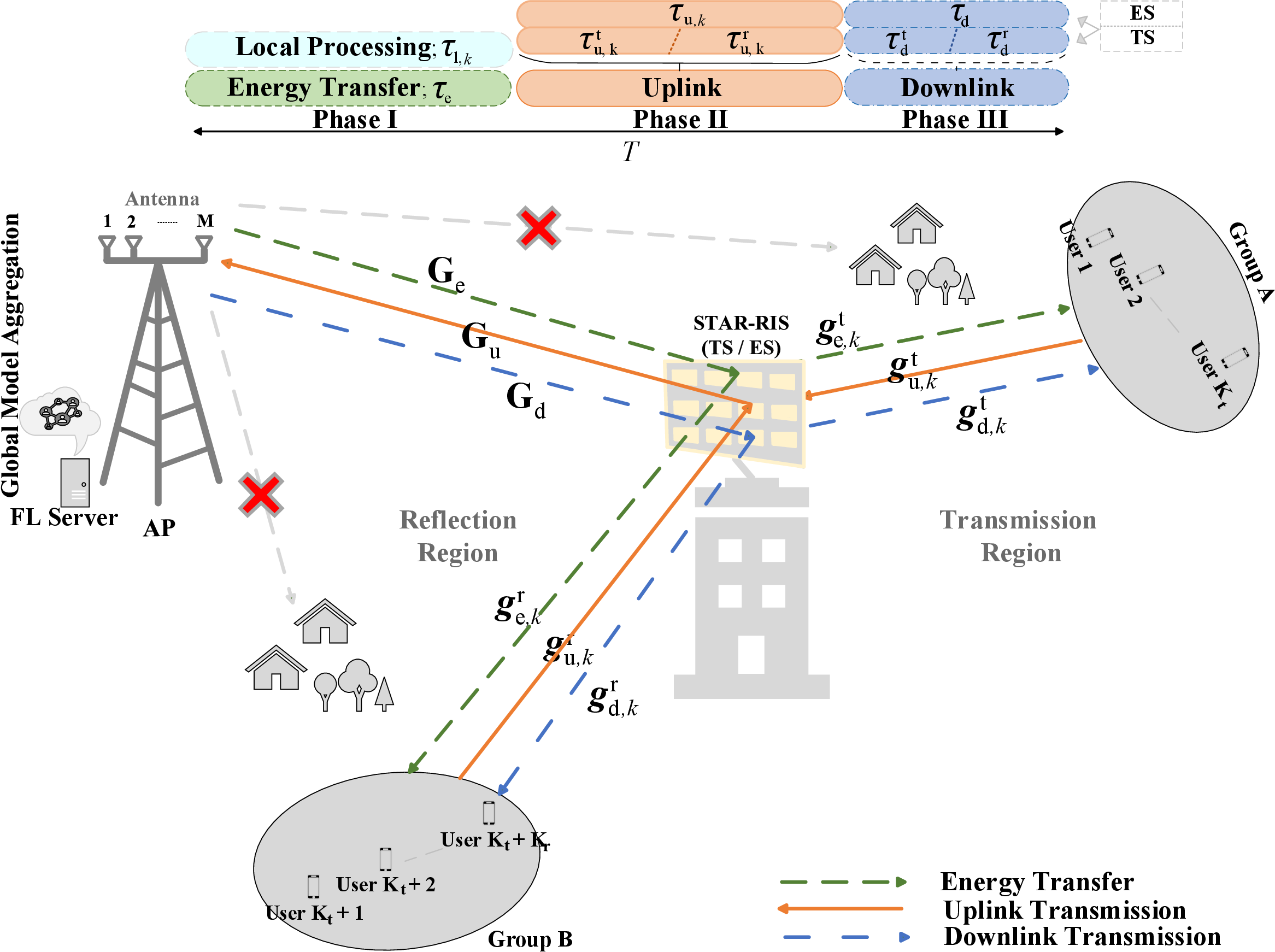}
  \caption{STAR-RIS aided WPT-FL multi-antenna AP system.}
  \label{fig:1}
\end{figure}

Figure \ref{fig:1} illustrates a STAR-RIS assisted WPT-FL multi-antenna AP network in three phases, namely energy transfer, uplink, and downlink transmission, denoted as $\mathcal{X} = \{\text{e}, \text{u}, \text{d}\}$\footnote{In this work, we optimize the allocation of all resources for one communication round of FL, consisting of three phases.}. The network consists of an AP equipped with $M$ antennas, indexed by $\mathcal{M} = \{1, \dots, M\}$; two distinct groups A and B with $K_{\text{t}}$ and $K_\text{r}$ single-antenna users, represented as $\mathcal{K}_\text{t} = \{1, \dots, K_\text{t}\}$ and $\mathcal{K}_\text{r} = \{K_\text{t} + 1, \dots, K_\text{t} + K_\text{r}\}$, respectively; a STAR-RIS with $N$ elements, indexed by $\mathcal{N} = \{1, \dots, N\}$. All users and the AP exchange information exclusively through the reflection and transmission spaces of the STAR-RIS, indicated by $\mathcal{Y} = \{\text{t}, \text{r}\}$, due to the absence of direct links.  The channel coefficients between the AP and STAR-RIS in all phases are denoted as $\textbf  G_{\mathcal{X}} \in \mathbb C^{N \times M}$, whereas $\textbf g^{\mathcal{Y}}_{\mathcal{X},k} \in \mathbb C^{N \times 1}$ is either the transmission or reflection channels gain from the STAR-RIS to device $k \in \mathcal{K}_\text{t} \cup \mathcal{K}_\text{r}$. In each phase, the beamforming matrix is applied to facilitate communication, denoted as $\mathbf{V}_{\mathcal{X}} = [\mathbf {v}_{\mathcal{X}, 1}, \mathbf {v}_{\mathcal{X}, 2}, ..., \mathbf {v}_{\mathcal{X}, (K_\text{t} + K_\text{r})}] \in \mathbb{C}^{M \times (K_\text{t} + K_\text{r})}$, with $\mathbf{v}_{{\mathcal{X}}, k} \in \mathbb{C}^{M \times 1}$ $\forall k \in \mathcal{K}_\text{t} \cup \mathcal{K}_\text{r}$. Since each user is located either in the reflection or the transmission region, $\textbf g^\text{t}_{\mathcal{X}, k} \forall k \in \mathcal{K}_\text{r}$, $\textbf g^\text{r}_{\mathcal{X}, k}  \forall k \in \mathcal{K}_\text{t}$ are all zero. The noise vector in the AP is modeled as additive white Gaussian noise with a covariance matrix \(\sigma_0 \mathbf{I}_M\), following the distribution \(\mathcal{C} \mathcal{N}(0, \sigma_0 \mathbf{I}_M)\), while the variance of additive white Gaussian noise for user $k$ is \(\sigma_k^2\) \footnote{ We assume perfect CSI across all phases, an ideal scenario that sets a lower bound for energy minimization in practice like \cite{10028982}.}.

Let us denote the amplitude coefficients and the corresponding phase shifter in transmission and reflection as $\alpha^{\mathcal{Y}}_{\mathcal{X},n}, ~\forall n \in \mathcal{N}$ and $\theta^{\mathcal{Y} }_{\mathcal{X},n} \in [0, 2\pi), ~\forall n \in \mathcal{N}$, the phase shift vectors of STAR-RIS across all phases can be defined as \cite{mu2021simultaneously}
\begin{equation}
    \textbf{$\boldsymbol \phi$}^{\mathcal{Y} }_{\mathcal{X}} = \bigg{(}\sqrt{\alpha^{\mathcal{Y} }_{\mathcal{X},1}}e^{j\theta^{\mathcal{Y} }_{\mathcal{X},1}},...,\sqrt{\alpha^{\mathcal{Y} }_{\mathcal{X},N}}e^{j\theta^{\mathcal{Y} }_{\mathcal{X},N}}\bigg{)}^\intercal
\end{equation}

By properly adjusting the amplitude coefficients and the corresponding phase shift for transmission and reflection, a given element of STAR-RIS can be operated in TS, ES, or MS modes \cite {mu2021simultaneously}. To be more specific, ES protocol simultaneously splits each element of STAR-RIS into two parts; one part transmits the incident signal to Group A (transmission region), while the other part reflects it towards Group B (reflection region), i.e., $\alpha^{  \text{t}}_{\mathcal{X},n} + \alpha^{ \text{r}}_{\mathcal{X},n} = 1$, $\alpha^{\mathcal{Y} }_{\mathcal{X},n} \in [0,1]$. In the TS protocol, all elements of STAR-RIS transmit the incident signal to either Group A or Group B in different time slots, i.e., $\alpha^{\text{t}}_{\mathcal{X},n} = 1$ and $\alpha^{\text{r}}_{\mathcal{X},n} = 0 $ in the first time slot, and $\alpha^{\text{t}}_{\mathcal{X},n} = 0$ and $\alpha^{\text{r}}_{\mathcal{X},n} = 1 $ in the second time slot\footnote{MS is a particular case of ES, i.e., $\alpha^{ \text{t}}_{\mathcal{X},n} + \alpha^{ \text{r}}_{\mathcal{X},n} = 1$, $\alpha^{\mathcal{Y}}_{\mathcal{X},n} \in \{0,1\}$, and since ES can be configured to match or outperform MS performance, this work ignores MS STAR-RIS \cite{mu2021simultaneously}.}.

With the phase shift vectors of STAR-RIS and beamforming matrices, the channels between the AP and user $k$ in energy transfer, uplink transmission, or downlink transmission can be represented as 
\begin{equation} \label{channel}
    z_{\mathcal{X},k} = \sum_{ \mathcal{Y} = \{\text{t}, \text{r}\}}|(\boldsymbol{\phi}^{\mathcal{Y}}_{\mathcal{X}} )^H \text{diag} (\textbf{g}^{\mathcal{Y}}_{\mathcal{X}, k}) \textbf{G}_{\mathcal{X}} \textbf{v}_{\mathcal{X}, k}|^2, \mathcal{X} \in \{\text{e}, \text{u}, \text{d}\}.
\end{equation}

To promote fairness among users, we impose the following proportional fairness constraint on the effective channel gains:
\begin{equation} \label{fairness}
    z_{\mathcal{X},i} = \beta_{ij} ~ z_{\mathcal{X},j},\quad \forall i,j \in \mathcal{K}_\text{t} \cup \mathcal{K}_\text{r},\ \mathcal{X} \in \{\text{e}, \text{u}, \text{d}\},
\end{equation}
where \( \beta_{ij} = \frac{\iota_i}{\iota_j} \) denotes the ratio of the Euclidean distances from the STAR-RIS to users \( i \) and \( j \). This constraint ensures that the effective channel gains are balanced proportionally based on user locations, thereby mitigating the double near-far effect commonly encountered in STAR-RIS-assisted systems.

We assume the STAR-RIS operates in ES mode for energy transfer, ensuring higher gain, flexibility, and fair energy harvesting for all users in transmission and reflection spaces, while consistently operating in TS or ES mode across uplink and downlink transmission \cite{10304608}.

\subsection{Phase I: Energy Transfer and Local Processing}
Users equipped with rechargeable batteries are ideally modeled with linear energy harvesting from the AP through STAR-RIS links over time $\tau_{\text{e}}$  \cite{10032506, 10028982}. The received energy for user $k$ is given by \cite{10032506, 10028982}
\begin{equation}
    E_k =   \eta P_{\text{e}} z_{\text{e},k} \tau_{\text{e}}, \forall k \in \mathcal{K}_\text{t} \cup \mathcal{K}_\text{r},
\end{equation}
where $P_{\text{e}}$ and $\eta$ $\in (0,1)$ represent the power of AP and energy conversion efficiency factor, respectively.

Utilizing harvested energy, each user processes its $L_{{\text{l}},k}$ raw local data at a computation frequency of $f_k$ within a time frame of $\tau_{{\text{l}},k}$. The local processing time is constrained by $\tau_{{\text{l}},k}\ge \tau_{\text{e}}$ to minimize its energy consumption  \cite{10032501}. To complete all required computations, i.e., $L_{{\text{l}},k}=\frac{f_k \tau_{{\text{l}},k}}{C_k}$, the local energy consumption is \cite{10032506, 10032501}
\begin{equation} \label{local}
E_{{\text{l}}, k} =   a{f_k}^3 \tau_{{\text{l}},k}, \forall k \in \mathcal{K}_\text{t} \cup \mathcal{K}_\text{r},
\end{equation}
where $a$ and $C_k$ are the energy coefficients depending on the chip architecture and computational complexity task at user $k$, respectively.  

\subsection{Phase II: Uplink Transmission}

Upon completion of local processing, all users simultaneously offload their model updates' data, each with the size of $L_{{\text{u}},k}$, to the AP under NOMA. Successive interference cancellation (SIC) is performed at the AP to mitigate the intra-cluster interference. Without loss of generality, it is assumed that the users are decoded following a descending indexing order of their channel gains \cite{6868214}. The achievable data rate and transmission time of each user during the uplink transmission depend on the STAR-RIS mode, as detailed below.

\subsubsection{ES mode}
Since all elements of STAR-RIS simultaneously serve both reflection and transmission regions in ES mode, each user encounters both intra-cluster and inter-cluster interference, given by
\begin{equation}
    \mathcal{J}^{\text{ES}}_{k} = \displaystyle \sum_{{m \notin \mathcal{I}^{'}_k, m>k}} |\sqrt{p_{\text{u}, m}} \textbf{h}_{\text{u},m} \textbf{v}_{\text{u}, k}|^2 + \sum_ {{m { \in} \mathcal{I}^{'}_k}} |\sqrt{p_{\text{u}, m}} \textbf{h}_{\text{u},m} \textbf{v}_{\text{u}, k}|^2,
\end{equation}
where $p_{\text{u}, m}$ is the power allocation to each user, $\textbf{h}_{\text{u},m} = \sum_{ \mathcal{Y} = \{\text{t}, \text{r}\}}(\boldsymbol{\phi}^{\mathcal{Y}}_{\text{u}} )^H \text{diag} (\textbf{g}^{\mathcal{Y}}_{\text{u}, m}) \textbf{G}_{\text{u}}$, and
\begin{equation}
\mathcal{I}^{'}_k =
  \begin{cases}
  \mathcal{K}_{\text{r}} ~~ \text{if} & k \in \mathcal{K}_{\text{t}}, \\
  \mathcal{K}_{\text{t}} ~~ \text{if} & k \in \mathcal{K}_{\text{r}}. \\
  \end{cases}
\end{equation}
Therefore, the achievable transmission rate and the corresponding delay for each user in transmitting $L_{{\text{u}},k}$ bits of data are respectively given by
\begin{equation} \label{R_u_ES}
\begin{aligned}
   R^{\text{ES}}_{\text{u}, k} =  \log_2 \bigg{(}1+\frac{p_{\text{u}, k} z_{\text{u},k}}{\mathcal{J}^{\text{ES}}_{k} + {|\sigma  \textbf{v}_{\text{u}, k}|^2}}\bigg{)},
\end{aligned}
\end{equation} 
\begin{equation} \label{t_k1}
\tau^{\text{ES}}_{{\text{u}},k}=\frac{L_{{\text{u}},k}}{R^{\text{ES}}_{\text{u}, k}}.
\end{equation}

\subsubsection{\text{TS} mode}
In TS mode, all elements serve either the reflection or transmission region, and thus, each user only faces intra-cluster interference, i.e.,
\begin{equation}
    \mathcal{J}^{\text{TS}}_{k} = \displaystyle \sum_{m \in \mathcal{I}_k , {m > k}} |\sqrt{p_{\text{u}, m}} \textbf{h}_{\text{u},m} \textbf{v}_{\text{u}, k}|^2,
\end{equation} 
where 
\begin{equation}
\mathcal{I}_k =
  \begin{cases}
  \mathcal{K}_{\text{t}} ~~ \text{if} & k \in \mathcal{K}_{\text{t}}, \\
  \mathcal{K}_{\text{r}} ~~ \text{if} & k \in \mathcal{K}_{\text{r}}. \\
  \end{cases}
\end{equation}
Then, the achievable transmission rate for each user and the corresponding time are respectively given by
\begin{equation} \label{R_u_TS}
\begin{aligned}
   R^{\text{TS}}_{\text{u}, k} =  \log_2 \bigg{(}1+\frac{p_{\text{u}, k} z_{\text{u},k}}{\mathcal{J}^{\text{TS}}_{k} + {|\sigma  \textbf{v}_{\text{u}, k}|^2}}\bigg{)},
\end{aligned}
\end{equation} 
\begin{equation} \label{t_k}
\tau^{\text{TS}}_{{\text{u}},k}=\frac{L_{{\text{u}},k}}{R^{\text{TS}}_{\text{u}, k}}.
\end{equation}

For any STAR-RIS operation mode, the energy consumed during processing and uploading for user $k$ must not exceed the energy harvested by that user; that is \cite{10032506, 10720523}, 
\begin{equation}
    E_{{\text{l}},k} + E^{\mathcal{S}}_{{\text{u}},k} \leq E_{k},~\forall k \in \mathcal{K}_{\text{t}} \cup \mathcal{K}_{\text{r}},  \mathcal{S} \in \{\text{TS}, \text{ES}\}, 
\end{equation}
where $E^{\mathcal{S}}_{{\text{u}},k} = p_{\text{u}, k}\tau^{\mathcal{S}}_{{\text{u}},k}$ is the energy consumption for user $k$ during uplink transmission.

\subsection{Phase III: Downlink Transmission}
The AP aggregates trained local models and then broadcasts the new global model with size $L_{\text{d}}$ back to all users. The downlink transmission rate and time are determined by the STAR-RIS operation mode as follows:

\subsubsection{ES mode}
All users receive the global model with power $P^{\text{ES}}_{\text{d}}$ and consequently, the downlink transmission rate and time are respectively given by
\begin{equation} \label{AP_power1}
R^{\text{ES}}_{\text{d}} = \log_2\bigg{(}1+{P^{\text{ES}}_{\text{d}}z_{\text{worst}}}\bigg{)}, 
\end{equation}
\begin{equation}
    \tau^{\text{ES}}_{\text{d}}=\frac{L_{\text{d}}}{R^{\text{ES}}_{\text{d}}},
\end{equation}
where $z_{\text{worst}}=  \underset{k}{\text{min}}\bigg{(}\frac{z_{\text{d},k}}{\sigma^2_k}\bigg{)}, \forall k \in \mathcal{K}_{\text{t}} \cup \mathcal{K}_{\text{r}}$ from \eqref{channel}. 

\subsubsection{TS mode}
Group A downloads the global model within a power of $P^{{\text{t}}}_{\text{d}}$ over full transmission, while Group B receives data with the power of $P^{{\text{r}}}_{\text{d}}$ during full reflection. The new global model is downloaded through the worst channel in each group, with the transmission time and rate given as follows.
\begin{equation} \label{AP_power}
R^{\text{TS}, \mathcal{Y}}_{\text{d}} = \log_2\bigg{(}1+{P^{\mathcal{Y}}_{\text{d}}z^{\mathcal{Y}}_{\text{worst}}}\bigg{)}, 
\end{equation}
\begin{equation} \label{AP_power1}
    \tau^\mathcal{Y}_{\text{d}}=\frac{L_{\text{d}}}{R^{\text{TS}, \mathcal{Y}}_{\text{d}}}, \mathcal{Y} \in \{\text{t}, \text{r}\},
\end{equation}
where
\begin{equation}
z^{\mathcal{Y}}_{\text{worst}}=
  \begin{cases}
 \underset{k}{\text{min}}\bigg{(}\frac{z_{\text{d},k}}{\sigma^2_k}\bigg{)} ~~ \text{if} & k \in \mathcal{K}_\text{t}, \mathcal{Y} = \text{t}, \\
  \underset{k}{\text{min}}\bigg{(}\frac{z_{\text{d},k}}{\sigma^2_k}\bigg{)} ~~ \text{if} & k \in \mathcal{K}_\text{r}, \mathcal{Y} = \text{r}. \\
  \end{cases}
\end{equation}

\section{Problems Formulation}
Considering different operation modes of STAR-RIS in uplink and downlink transmissions (e.g., ES-ES indicates ES mode for STAR-RIS in both uplink and downlink transmissions), we formulate the following energy minimization problems for the multi-antenna AP STAR-RIS aided WPT-FL network, each under the overall delay constraint \( T \).

\subsection{ES-ES}
Assuming ES STAR-RIS for both transmission phases, i.e., ES-ES, the formulated energy minimization problem can be expressed by
\begin{subequations} \label{main_problem-ES_ES}
\begin{align}
\min_{\mathcal{V}_1} \quad & P_{\text{e}} \tau_{\text{e}} + P^{\text{ES}}_{\text{d}} \tau^{\text{ES}}_{\text{d}}, \\&\ \hspace{-1.4cm} \text{s.t.}~ \tau_{{\text{l}},k} + \tau^{\text{ES}}_{{\text{u}},k} + \tau^{\text{ES}}_{\text{d}} \leq  T,~\forall k \in \mathcal{K}_\text{t} \cup \mathcal{K}_\text{r},\\
& \hspace{-0.8cm} \tau_{{\text{l}},k} \ge \tau_{\text{e}}, ~\forall k \in \mathcal{K}_\text{t} \cup \mathcal{K}_\text{r}, \\
& \hspace{-0.8cm} \eta P_{\text{e}} z_{\text{e},k} \tau_{\text{e}} \geq p_{\text{u}, k} \tau^{\text{ES}}_{{\text{u}},k} + af^3_{k} \tau_{{\text{l}},k}, ~\forall k \in \mathcal{K}_\text{t} \cup \mathcal{K}_\text{r}, \\
&\hspace{-0.8cm} L_{{\text{l}},k} = \frac{f_k \tau_{{\text{l}},k}}{C_k},~ \forall k \in \mathcal{K}_\text{t} \cup \mathcal{K}_\text{r}, \\
&\hspace{-0.8cm} L_{{\text{u}},k} = R^{\text{ES}}_{\text{u}, k} \tau^{\text{ES}}_{{\text{u}},k}, ~\forall k \in \mathcal{K}_\text{t} \cup \mathcal{K}_\text{r}, \\
&\hspace{-0.8cm} L_{\text{d}} = R^{\text{ES}}_{\text{d}} \tau^{\text{ES}}_{\text{d}}, \\
&\hspace{-0.8cm} |\textbf{$\phi$}^{ \text{t}}_{\mathcal{X},n}|^2 + |\textbf{$\phi$}^{ \text{r}}_{\mathcal{X},n}|^2 = 1, ~\forall n \in \mathcal{N},~ \mathcal{X} = \{\text{e}, \text{u}, \text{d}\}, \\
&\hspace{-0.8cm} ||\textbf{V}_{\mathcal{X}}||_F^2 \leq 1,~ \mathcal{X} = \{\text{e}, \text{u}, \text{d}\}, \\
&\hspace{-0.8cm} z_{\mathcal{X},i} = \mathbf{\beta}_{ij} \times  z_{\mathcal{X},j},~\forall i,j \in \mathcal{K}_\text{t} \cup \mathcal{K}_\text{r}, \mathcal{X} = \{\text{e}, \text{u}, \text{d}\}, \\
&\hspace{-0.8cm} P_{\text{e}} \leq P_{\text{max}}, \\
&\hspace{-0.8cm} p_{\text{u}, k} \leq p^{\text{max}}_{k}, ~\forall k \in \mathcal{K}_\text{t} \cup \mathcal{K}_\text{r}, \\
&\hspace{-0.8cm} P^{\text{ES}}_{\text{d}} \leq P_{\text{max}},
\end{align}
\end{subequations}
where $\mathcal{V}_1 = \{ \tau_{\text{e}},\tau_{{\text{l}},k},\tau^{\text{ES}}_{{\text{u}},k}, \tau^{\text{ES}}_{\text{d}}, P_{\text{e}}, p_{\text{u}, k}, P^{\text{ES}}_{\text{d}}, f_k,\textbf{V}_{\mathcal{X}}, \textbf{$\boldsymbol \phi$}^{\mathcal{Y}}_{\mathcal{X}}\}$,  constraint in (\ref{main_problem-ES_ES}\rm{b}) limits system latency; (\ref{main_problem-ES_ES}\rm{c}) indicates that no uploading is performed while the AP transfers power, aiming to save energy; (\ref{main_problem-ES_ES}\rm{d}) limits the total consumed energy of each user by the harvested energy of that user; Constraints~(\ref{main_problem-ES_ES}\rm{e}) and~(\ref{main_problem-ES_ES}\rm{f}) specify the required data sizes to be locally processed and offloaded for each user, respectively, whereas constraint~(\ref{main_problem-ES_ES}\rm{g}) corresponds to the data size that must be downloaded from the AP; (\ref{main_problem-ES_ES}\rm{h}) constrains phase shifts and amplitudes of the STAR-RIS for all phases; (\ref{main_problem-ES_ES}\rm{i}) constrains the energy transfer, active, and downlink beamformers’ gain for the AP; (\ref{main_problem-ES_ES}\rm{j}) ensures quality of service by promoting fairness both across user groups and within individual groups; (\ref{main_problem-ES_ES}\rm{k}) and (\ref{main_problem-ES_ES}\rm{m}) restrict the power of AP over-harvesting and downloading phases, while (\ref{main_problem-ES_ES}\rm{l}) limits the maximum power allocation to each user with $p^{\text{max}}_{k}$.

\subsection{ES-TS}
Suppose that STAR-RIS operates in ES and TS modes during uplink and downlink transmission, respectively. The energy minimization problem can be formulated as 
\begin{subequations} \label{main_problem-ES_TS}
\begin{align}
\min_{\mathcal{V}_2} \quad & P_{\text{e}} \tau_{\text{e}} + \sum_{\mathcal{Y} =\{\text{t}, \text{r}\}} P^{\mathcal{Y}}_{\text{d}} \tau^{\mathcal{Y}}_{\text{d}}, \\&\ \hspace{-1.3cm} \text{s.t.}~ \tau_{{\text{l}},k} + \tau^{\text{ES}}_{{\text{u}},k} + \sum_{\mathcal{Y} =\{\text{t}, \text{r}\}} \tau^{\mathcal{Y}}_{\text{d}} \leq  T,~~\forall k \in \mathcal{K}_\text{t} \cup \mathcal{K}_\text{r},\\
&\hspace{-0.8cm} L_{\text{d}} = R^{\text{TS}, \mathcal{Y}}_{\text{d}} \tau^{\mathcal{Y}}_{\text{d}}, ~ \mathcal{Y} = \{\text{t}, \text{r}\} \\
&\hspace{-0.8cm} |\textbf{$\phi$}^{ \text{t}}_{\mathcal{X}_{2},n}|^2 + |\textbf{$\phi$}^{ \text{r}}_{\mathcal{X}_{2},n}|^2 = 1, ~\forall n \in \mathcal{N},~ \mathcal{X}_{2} = \{\text{e}, \text{u}\}, \\
&\hspace{-0.8cm} |\textbf{$\phi$}^{ \text{t}}_{\text{d},n}| =1, |\textbf{$\phi$}^{ \text{r}}_{\text{d},n}| = 1,~\forall n \in \mathcal{N}, \\
&\hspace{-0.8cm} \sum_{\mathcal{Y} =\{\text{t}, \text{r}\}} P^{\mathcal{Y}}_{\text{d}} \leq P_{\text{max}},\\
&\hspace{-0.8cm}(\ref{main_problem-ES_ES}\rm{c} - \ref{main_problem-ES_ES}\rm{f}), (\ref{main_problem-ES_ES}\rm{i} - \ref{main_problem-ES_ES}\rm{l})
\end{align}
\end{subequations}
where $\mathcal{V}_2 = \{ \tau_{\text{e}},\tau_{{\text{l}},k},\tau^{\text{ES}}_{{\text{u}},k}, \tau^{\mathcal{Y}}_{\text{d}}, P_{\text{e}}, p_{\text{u}, k}, P^{\mathcal{Y}}_{\text{d}}, f_k,\textbf{V}_{\mathcal{X}}, \textbf{$\boldsymbol \phi$}^{\mathcal{Y}}_{\mathcal{X}}\}$. Compared to problem \eqref{main_problem-ES_ES}, the objective function, the constraints on overall delay, downlink data rate, and the STAR-RIS amplitude for downlink transmission have changed for problem \eqref{main_problem-ES_TS}.

\subsection{TS-ES}
In this scenario, users offload data through TS STAR-RIS, whereas the global model is downloaded under ES STAR-RIS. 
\begin{subequations} \label{main_problem-TS_ES}
\begin{align}
\min_{\mathcal{V}_3} \quad & P_{\text{e}} \tau_{\text{e}} +  P^{\text{ES}}_{\text{d}} 
\tau^{\text{ES}}_{\text{d}},
\\&\ \hspace{-1.3cm} \begin{aligned}
\text{s.t.}~ \tau_{{\text{l}},k} +  \tau^{\text{TS}}_{{\text{u}},k} + \tau^{\text{TS}}_{{\text{u}},k^{'}} + \tau^{\text{ES}}_{\text{d}} \leq  T,~~~ & \substack{\text{\normalsize $\forall k \in \mathcal{K}_\text{t} \cup \mathcal{K}_\text{r},$} \\[3pt] \text{\normalsize $k^{'} \in \mathcal{I}^{'}_k,$}}
\end{aligned} \\
& \hspace{-0.6cm} \eta P_{\text{e}} z_{\text{e},k} \tau_{\text{e}} \geq  p_{\text{u}, k} \tau^{\text{TS}}_{{\text{u}},k} + af^3_{k} \tau_{{\text{l}},k}, ~\forall k \in \mathcal{K}_\text{t} \cup \mathcal{K}_\text{r}, \\
&\hspace{-0.6cm} L_{\text{u}, k} = R^{\text{TS}}_{\text{u}, k} \tau^{\text{TS}}_{\text{u}, k}, \forall k \in \mathcal{K}_\text{t} \cup \mathcal{K}_\text{r}, \\
&\hspace{-0.6cm} |\textbf{$\phi$}^{ \text{t}}_{\mathcal{X}_{3},n}|^2 + |\textbf{$\phi$}^{ \text{r}}_{\mathcal{X}_{3},n}|^2 = 1, ~\forall n \in \mathcal{N},~ \mathcal{X}_{3} = \{\text{e}, \text{d}\}, \\
&\hspace{-0.6cm} |\textbf{$\phi$}^{ \text{t}}_{\text{u},n}| =1, |\textbf{$\phi$}^{ \text{r}}_{\text{u},n}| = 1,~\forall n \in \mathcal{N}, \\
&\hspace{-0.6cm} (\ref{main_problem-ES_ES}\rm{c}), (\ref{main_problem-ES_ES}\rm{e}), (\ref{main_problem-ES_ES}\rm{g}), (\ref{main_problem-ES_ES}\rm{i} - \ref{main_problem-ES_ES}\rm{m}),
\end{align}
\end{subequations}
where $\mathcal{V}_3 = \{ \tau_{\text{e}},\tau_{{\text{l}},k},\tau^{\text{TS}}_{{\text{u}},k}, \tau^{\text{ES}}_{\text{d}}, P_{\text{e}}, p_{\text{u}, k}, P^{\text{ES}}_{\text{d}}, f_k,\textbf{V}_{\mathcal{X}}, \textbf{$\boldsymbol \phi$}^{\mathcal{Y}}_{\mathcal{X}}\}$. The constraints on overall delay, uplink data rate, and STAR-RIS amplitudes in uplink transmission differ from the problem \eqref{main_problem-ES_ES}.

\subsection{\text{TS}-\text{TS}}
In both uplink and downlink transmission, STAR-RIS operates in TS, and thus, the energy minimization problem is represented by 
\begin{subequations} \label{main_problem-TS_TS}
\begin{align}
\min_{\mathcal{V}_4} \quad & P_{\text{e}} \tau_{\text{e}} + \sum_{\mathcal{Y} =\{\text{t}, \text{r}\}} P^{\mathcal{Y}}_{\text{d}} \tau^{\mathcal{Y}}_{\text{d}},
\\&\ \hspace{-1.3cm} \begin{aligned}
\text{s.t.}~ \tau_{{\text{l}},k} +  \tau^{\text{TS}}_{{\text{u}},k} + \tau^{\text{TS}}_{{\text{u}},k^{'}} + \sum_{\mathcal{Y} =\{\text{t}, \text{r}\}} \tau^{\mathcal{Y}}_{\text{d}} \leq  T, & \substack{\text{\normalsize $\forall k \in \mathcal{K}_\text{t} \cup \mathcal{K}_\text{r},$} \\[3pt] \text{\normalsize $k^{'} \in \mathcal{I}^{'}_k,$}}
\end{aligned} \\
&\hspace{-0.6cm} |\textbf{$\phi$}^{ \text{t}}_{\text{e},n}|^2 + |\textbf{$\phi$}^{ \text{r}}_{\text{e},n}|^2 = 1, ~\forall n \in \mathcal{N}, \\
& \hspace{-0.7cm} \begin{aligned}
& \substack{\text{\normalsize (\ref{main_problem-ES_ES}\rm{c}), (\ref{main_problem-ES_ES}\rm{e}),(\ref{main_problem-ES_ES}\rm{i} - \ref{main_problem-ES_ES}\rm{l}), (\ref{main_problem-ES_TS}\rm{c}),} \\[3pt] \text{\normalsize (\ref{main_problem-ES_TS}\rm{e} - \ref{main_problem-ES_TS}\rm{f}), (\ref{main_problem-TS_ES}\rm{c} - \ref{main_problem-TS_ES}\rm{d}), (\ref{main_problem-TS_ES}\rm{f}),}}
\end{aligned} 
 \end{align}
\end{subequations}
where $\mathcal{V}_4 = \{ \tau_{\text{e}},\tau_{{\text{l}},k},\tau^{\text{TS}}_{{\text{u}},k}, \tau^{\mathcal{Y}}_{\text{d}}, P_{\text{e}}, p_{\text{u}, k}, P^{\mathcal{Y}}_{\text{d}}, f_k,\textbf{V}_{\mathcal{X}}, \textbf{$\boldsymbol \phi$}^{\mathcal{Y}}_{\mathcal{X}}\}$. The objective function and the constraints on the overall delay, data rate, and STAR-RIS amplitudes for uplink and downlink transmissions differ from problem \eqref{main_problem-ES_ES}.

\section{Proposed Solution}

In each scenario, the formulated optimization problem is non-convex due to the non-convexity of the objective function, as well as constraints on total energy consumption for each user, data rates in both uplink and downlink transmission, and the phase shift vectors of the STAR-RIS. Solving each problem directly is challenging, and thus, we decouple it into two blocks. Figure 2 presents a concise flowchart of the proposed method for all scenarios, where Block B jointly optimizes time, power across all phases, and computation frequency for each user, while Block A independently optimizes phase shift vectors of the STAR-RIS and beamforming matrices.
\begin{figure}[hbpt]
\hspace{-0.5cm}
\centering
\begin{tikzpicture}[thick, scale=0.25, every node/.style={scale=0.75}]
    \node[ellipse, draw, shape aspect=4, text width=16em, minimum height=4em] 
        at (-2.5, 14.3) (block1) {\begin{tabular}{c} \hspace{-0.6cm}\textbf{Problem Formulation} \\ $(\tau_{\text{e}}, \tau_{{\text{l}},k}, \tau_{{\text{u}},k}, \tau_{\text{d}}, f_{k}, P_{\text{e}}, P_{\text{d}}, p_{\text{u}, k}, \phi^{\mathcal{Y}}_{\mathcal{X}}, \textbf{V}_{\mathcal{X}})$ \end{tabular} };

    \node[ellipse, green!50!black!50!, draw, shape aspect=4, text width=6em, minimum height=3em, align=center] 
        at (-14.5, 7.7) (block2)  
        {\begin{tabular}{c} \textbf{Optimize $\phi^{\mathcal{Y}}_{\text{e}}$} \\ \textbf{and $\textbf{V}_{\text{e}}$} \end{tabular}};

        \node[ellipse,  dotted, green!50!black!70!, draw, shape aspect=4, text width=6em, minimum height=0.1em, align=center] 
        at (-14.5, 3.9) (block20)  
        {\begin{tabular}{c} \hspace{-0.3cm} \textbf{BCD $\rightarrow$ SDP} \end{tabular}};

    \node[ellipse, orange, draw, shape aspect=4, text width=6em, minimum height=3em, align=center] 
        at (-2.5, 7.7) (block3)  
        {\begin{tabular}{c} \textbf{Optimize $\phi^{\mathcal{Y}}_{\text{u}}$} \\ \textbf{and $\textbf{V}_{\text{u}}$} \end{tabular}};

         \node[ellipse,  dotted, orange, draw, shape aspect=4, text width=6em, minimum height=0.1em, align=center] 
        at (-2.5, 3.9) (block30)  
        {\begin{tabular}{c} \hspace{-0.5cm} \textbf{BCD $\rightarrow$ SDP/RQ} \end{tabular}};

    \node[ellipse, blue!70, draw, shape aspect=4, text width=6em, minimum height=3em, align=center] 
        at (9.5, 7.7) (block4)  
        {\begin{tabular}{c} \textbf{Optimize $\phi^{\mathcal{Y}}_{\text{d}}$} \\ \textbf{and $\textbf{V}_{\text{d}}$} \end{tabular}};

        \node[ellipse,  dotted, blue, draw, shape aspect=4, text width=6.3em, minimum height=0.1em, align=center] 
        at (9.5, 3.9) (block40)  
        {\begin{tabular}{c} \hspace{-0.1cm}\textbf{BCD $\rightarrow$ SDP} \end{tabular}};

        \node[draw=none,  minimum width=0.05cm, minimum height=0.1cm, ->, rotate=0, fill=white] at (-17.4, 11.4) (block152) {\bf Block A};

        \node[draw=none,  minimum width=0.05cm, minimum height=0.1cm, ->, rotate=0, fill=white] at (-8, 1) (block152) {\bf Block B};

    \node[ellipse, brown,draw, shape aspect=2, text width=12em, minimum height=4em, align=center, text centered] 
        at (-2.5, -1.5) (block13)  
        {\textbf{Optimize $\tau_{\text{e}}, \tau_{{\text{l}},k}, \tau_{{\text{u}},k}, \tau_{\text{d}}, f_{k}, P_{\text{e}}, p_{\text{u}, k}, P_{\text{d}}$}};

        \node[ellipse,  dotted, brown, draw, shape aspect=4, text width=6em, minimum height=0.1em, align=center] 
        at (-15.9, -1.5) (block50)  
        {\begin{tabular}{c} \textbf{ODSAs / BA} \end{tabular}};

    \node[ellipse, draw, shape aspect=4, text width=6em, minimum height=4em] 
        at (-2.5, -7.5) (block14)  
        {\begin{tabular}{c} \hspace{-0.3cm}\textbf{  $P^{*}_{\text{e}} \tau_{\text{e}}^* + P^{*}_{\text{d}}\tau_{\text{d}}^*$ } \end{tabular}};

    \node[draw, gray!50!, minimum width=33.5em, minimum height=8em, ->] 
        at (-2.5, 6.4) (block16) {};
        
    \node[draw, dotted, black!100!, minimum width=10em, minimum height=7.3em, ->] 
        at (-14.5, 6.4) (block161) {};
        
    \node[draw, dotted, black!100!, minimum width=10em, minimum height=7.3em, ->] 
        at (-2.5, 6.4) (block162) {};

    \node[draw, dotted, black!100!, minimum width=10em, minimum height=7.3em, ->] 
        at (9.4, 6.4) (block163) {};

     


    \draw[-latex, line width=2pt, gray, ->] (block16) edge (block13);
    \draw[-latex, line width=2pt, gray, ->] (block13) edge (block14);
    \draw[thick, line width=2pt, gray, ->] (block1.south) -- ++(0,0)  -| (block16.north);
\end{tikzpicture}
\caption{Flowchart of the proposed optimization method across all scenarios.}
\label{fig2}
\end{figure}
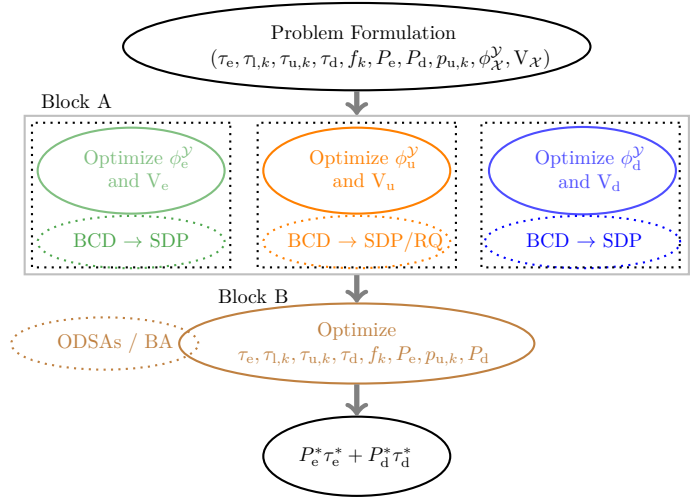

To simplify the problem in each scenario, we first derive the optimized power of AP in energy transfer in the following theorem.
\begin{theorem} \label{theorem1}
    The optimal $P_\text{e}$ equals $P_{\max}$. 
\end{theorem}

\begin{IEEEproof} By reducing \( \tau_{\text{e}} \), more time is freed up for local processing, uplink transmission, and downlink transmission, enabling more efficient resource allocation and reducing total energy consumption. Thus, minimizing \( \tau_{\text{e}} \) is the best option under equal harvested energy.  In the following, we prove that setting \( P_{\text{max}} \) as the AP’s optimal power in the transfer phase not only minimizes \( \tau_{\text{e}} \), but also consumes no more energy than any \( P_{\text{e}} \in (0, P_{\text{max}}) \) in the energy transfer phase.

Assume the optimal power of AP in energy transfer is $P_{\text{e}}^{*} < P_{\text{max}}$ with harvesting time $\tau^{*}_{\text{e}}$. If we set $P_{\text{e}} = P_{\text{max}}$, the same energy can be harvested in a shorter time, i.e., $\tau_{\text{e}} < \tau_{\text{e}}^{*}$, relaxing constraints (\ref{main_problem-ES_ES}\rm{b})-(\ref{main_problem-ES_ES}\rm{d}). This results in a better solution, contradicting the assumption that $P_{\text{e}}^{*}$ is optimal. Thus, the optimal power must be $P_{\text{max}}$ \cite{10400811}.
\end{IEEEproof}

\subsection{Optimization of Phase Shift Vectors of STAR-RIS and Beamforming Matrices in All Phases; $\textbf{$\phi$}^{\mathcal{Y}}_{\mathcal{X}}$ and $\textbf{V}_{\mathcal{X}}$}
The phase shift vectors of the STAR-RIS and the beamforming matrix for each phase can be optimized independently through the following optimization problems. Before all that, we define ${\textbf{V}}_{\mathcal{X}, k} = \textbf{v}_{\mathcal{X}, k} \textbf{v}^H_{\mathcal{X}, k}$, $\Phi_{\mathcal{X}}^{\mathcal{Y}} = \boldsymbol{\phi}^{\mathcal{Y}}_{\mathcal{X}} (\boldsymbol{\phi}^{\mathcal{Y}}_{\mathcal{X}})^H $, $\Lambda^{\mathcal{Y}}_{\mathcal{X}, k} = \text{diag} (\textbf{g}^{\mathcal{Y}}_{\mathcal{X}, k}) \textbf{G}_{\mathcal{X}} \textbf{V}_{\mathcal{X}, k} \textbf{G}^{H}_{\mathcal{X}} \text{diag} (\textbf{g}^{\mathcal{Y}}_{\mathcal{X}, k})^H$, and $\Gamma^{\mathcal{Y}}_{\mathcal{X}, k} = \textbf{G}^{H}_{\mathcal{X}} \text{diag} (\textbf{g}^{\mathcal{Y}}_{\mathcal{X}, k})^H \Phi_{\mathcal{X}}^{\mathcal{Y}} \text{diag} (\textbf{g}^{\mathcal{Y}}_{\mathcal{X}, k}) \textbf{G}_{\mathcal{X}}$$, \forall k \in \mathcal{K}_{\text{t}} \cup \mathcal{K}_{\text{r}}, \mathcal{X} =   \{\text{e},\text{u},\text{d}\} $, $ \mathcal{Y} = \{\text{t},\text{r}\} $.

\vspace{0.25 cm}
\subsubsection{Joint Optimization of Energy Transfer Phase Shift Vectors of STAR-RIS and Beamforming Matrix; $\textbf{$\phi$}^{\mathcal{Y}}_{\text{e}}$ and $\textbf{V}_{\text{e}}$}

Based on Theorem \ref{theorem1} and (\ref{main_problem-ES_ES}\rm{d}), maximizing the energy harvesting channel for each user minimizes the harvesting time. Since all users share the same phase shift vectors of the STAR-RIS and the beamforming matrix and must satisfy the constraint (\ref{main_problem-ES_ES}\rm{j}) for all users simultaneously, the following optimization problems are formulated to heuristically adjust the phase shift vectors of STAR-RIS and the beamforming matrix during the energy transfer phase. 
\begin{subequations} {\label{phaseenmain}}
\begin{align}
\max_{\textbf{V}_{\text{e}},\phi^{\mathcal{Y}}_{\text{e}}}~~\sum_{ k = \mathcal{K}_{\text{t}} \cup \mathcal{K}_{\text{r}}} z_{\text{e},k},\\
&\ \hspace{-2.8cm} \text{s.t.}~ |\textbf {$\phi$}^{\text{t}}_{\text{e},n}|^2 + |\textbf {$\phi$}^{\text{r}}_{\text{e},n}|^2 = 1, ~\forall n\in \mathcal{N},\\
&\ \hspace{-2.25cm} ||\textbf{V}_\text{e}||_F^2\leq 1,\\
&\hspace{-2.15cm} z_{\text{e},i} = \mathbf{\beta}_{ij} \times  z_{\text{e},j},~\forall i,j \in \mathcal{K}_\text{t} \cup \mathcal{K}_\text{r}.
\end{align}
\end{subequations}
The BCD algorithm is employed to iteratively update the STAR-RIS phase shift vectors and the beamforming matrix, terminating once the objective function of \eqref{phaseenmain} no longer improves. We first randomly initialize the phase shift vectors of the STAR-RIS and the active beamforming matrix such that they satisfy (\ref{main_problem-ES_ES}\rm{h}) and (\ref{main_problem-ES_ES}\rm{i}) in the energy transfer phase, where the phase shifters are in \( [0, 2\pi) \) for all elements in both transmission and reflection sides.

{\bf{ Update $\phi^{\mathcal{Y}}_{\text{e}}$:}} Under given $\textbf{V}^*_{\text{e}}$, the original problem can be simplified to maximizing \( z_{\text{e},k'} \) for an arbitrary user, say \( k' \), as the other \( z_{\text{e},k} \) are directly proportional to \( z_{\text{e},k'} \) from (\ref{phaseenmain}\rm{e}). Therefore, considering trace properties, \eqref{phaseenmain} can be reformulated as
\begin{subequations} {\label{etchannel}}
\begin{flalign}
\max_{ \Phi_{\text{e}}^{\mathcal{Y}}}~~~ \sum_{ \mathcal{Y} = \{\text{t}, \text{r}\}} {\text{Tr}}\bigg{(}  \Phi_{\text{e}}^{\mathcal{Y}} \Lambda^{\mathcal{Y}}_{\text{e}, k'} \bigg{)}, &&\\
&\ \hspace{-4.1cm}\text{s.t.}~{\Phi}^{\mathcal{Y}}_{\text{e}} \succeq 0,~ \mathcal{Y} = \{\text{t}, \text{r}\}, &&\\
& \hspace{-3.4cm} \text{diag}({{\Phi}^{\text{t}}_{\text{e}}}) + \text{diag}({{\Phi}^{\text{r}}_{\text{e}}})=1^{\it{N}}, &&\\
& \hspace{-3.4cm} {\text{Rank}} ({\Phi}^{\mathcal{Y}}_{\text{e}}) =1,~ \mathcal{Y} = \{\text{t}, \text{r}\},&&\\
&\ \hspace{-3.5cm} \frac{\sum_{ \mathcal{Y} = \{\text{t}, \text{r}\}}{\text{Tr}}\bigg{(}  \Phi_{\text{e}}^{\mathcal{Y}} \Lambda^{\mathcal{Y}}_{\text{e}, i} \bigg{)}}{\sum_{ \mathcal{Y} = \{\text{t}, \text{r}\}}{\text{Tr}}\bigg{(} \Phi_{\text{e}}^{\mathcal{Y}} \Lambda^{\mathcal{Y}}_{\text{e}, j} \bigg{)}} = \beta_{ij}, \forall i,j \in \mathcal{K}_\text{t} \cup \mathcal{K}_\text{r}. &&
\end{flalign}
\end{subequations} 
Removing the rank-1 constraint results in a convex SDP that can be solved using CVX.

{\bf{ Update $\textbf{V}_{\text{e}}$:}} Due to the coupling effect in the active beamforming matrix for all users, \eqref{phaseenmain} cannot be simplified to maximize a single user. Therefore, given $\phi^{\mathcal{Y}*}_{\text{e}}$, the corresponding beamforming matrix in the harvesting phase can be optimized as follows:
\begin{subequations} {\label{phaseen1}}
\begin{align}
\max_{\textbf{V}_{\text{e}}}~~\sum_{ k = \mathcal{K}_{\text{t}} \cup \mathcal{K}_{\text{r}}} z_{\text{e},k},\\
&\ \hspace{-2.8cm} \text{s.t.}~ z_{\text{e},i} = \beta_{ij} \times z_{\text{e},j}, ~\forall i,j \in \mathcal{K}_\text{t} \cup \mathcal{K}_\text{r},\\
&\hspace{-2.15cm} ||\textbf{V}_\text{e}||_F^2\leq 1.
\end{align}
\end{subequations}
Likewise, the problem can be reformulated as a SDP problem after dropping the rank-1 constraint.
\begin{subequations} {\label{etchannel1}}
\begin{flalign}
\max_{{\textbf{V}}_{\text{e}, k}}~~~\sum_{ k = \mathcal{K}_{\text{t}} \cup \mathcal{K}_{\text{r}}} \sum_{ \mathcal{Y} = \{\text{t}, \text{r}\}} {\text{Tr}}\bigg{(} \Gamma^{\mathcal{Y}}_{\text{e}, k} {\textbf{V}}_{\text{e}, k} \bigg{)}, &&\\
&\ \hspace{-5.1cm}\text{s.t.}~ {\textbf{V}}_{\text{e}, k} \succeq 0, ~\forall k \in \mathcal{K}_{\text{t}} \cup \mathcal{K}_{\text{r}}, &&\\
& \hspace{-4.5cm} \sum_{k = 1}^{K_{\text{t}} + K_{\text{r}}} {\text{Tr}} ( {\textbf{V}}_{\text{e}, k}) \leq 1, &&\\
&\ \hspace{-4.5cm} {\text{Rank}} ({\textbf{V}}_{\text{e}, k}) =1, ~\forall k \in \mathcal{K}_{\text{t}} \cup \mathcal{K}_{\text{r}}, &&\\
&\ \hspace{-4.5cm} \frac{\sum_{ \mathcal{Y} = \{\text{t}, \text{r}\}} {\text{Tr}}\bigg{(} \Gamma^{\mathcal{Y}}_{\text{e}, i} {\textbf{V}}_{\text{e}, i} \bigg{)}}{\sum_{ \mathcal{Y} = \{\text{t}, \text{r}\}} {\text{Tr}}\bigg{(} \Gamma^{\mathcal{Y}}_{\text{e}, j} {\textbf{V}}_{\text{e}, j} \bigg{)}} = \beta_{ij}, ~\forall i,j \in \mathcal{K}_\text{t} \cup \mathcal{K}_\text{r}. &&
\end{flalign}
\end{subequations}
Once the BCD is terminated, Gaussian randomization addresses the rank-1 constraint for both phase shift vectors of STAR-RIS and the beamforming matrix.

\subsubsection{Joint Optimization of Uplink Transmission Phase Shift Vectors of STAR-RIS, and Beamforming Matrix; $\textbf{$\phi$}^{\mathcal{Y}}_{\text{u}}$ and $\textbf{V}_{\text{u}}$}

According to Eq. (8) in \cite{Zeng_GC2018}, the energy consumption for each user in uplink transmission with TS STAR-RIS can be rewritten as follows:
\begin{equation} \label{equivalent}
\begin{split}
    p_{\text{u}, k} \tau^{\text{TS}}_{\text{u}, k} &= \\
    &\hspace{-1cm}\frac{
        2^{\tiny \bigg( \sum\nolimits_{m \notin \mathcal{I}^{'}_k ,\, m > k} R_{\text{u},m} \bigg)}  \tau^{\text{TS}}_{\text{u}, k} \left( 2^{R_{\text{u},k}} - 1 \right)  \left| \sigma \textbf{v}_{\text{u}, k} \right|^2
    }{
        z^{*}_{\text{u}, k}
    }, 
    \quad \forall k \in \mathcal{K}_\text{t} \cup \mathcal{K}_\text{r}.
\end{split}
\end{equation}
Through derivative analysis, it can be shown that $p_{{\text{u}, k}} \tau^{\text{TS}}_{\text{u}, k}, \forall k$ decreases with $\tau^{\text{TS}}_{\text{u}, k}$ \cite{Zeng_GC2018, alishahi2023latency}. Since lower energy consumption in uplink transmission reduces the required harvested energy per user, maximizing $\frac{z_{\text{u}, k}}{|\sigma  \textbf{v}_{\text{u}, k}|^2}$ for each user results in optimized phase shift vectors of STAR-RIS and the active beamforming matrix for uplink transmission. Note that this derivation also applies to the ES STAR-RIS in uplink transmission. Hence, we formulate the following optimization problem for each scenario to optimize the phase shift vectors of STAR-RIS and the active beamforming matrix. 

\paragraph{\text{ES}}
Considering NOMA in the uplink transmission, where all users share the same phase shift vectors of the STAR-RIS and the coupling effect in the active beamforming vectors for all users, the optimized phase shift vectors and beamforming matrix for uplink transmission can be heuristically obtained from (\ref{main_problem-ES_ES}) as follows:
\begin{subequations} {\label{phaseut}}
\begin{align}
\max_{\textbf{V}_{\text{u}},\phi^{\mathcal{Y}}_{\text{u}}}~~\sum_{ k \in \mathcal{K}_{\text{t}} \cup \mathcal{K}_{\text{r}}} \frac{z_{\text{u}, k}}{|\sigma  \textbf{v}_{\text{u}, k}|^2},\\
&\ \hspace{-2.8cm} \text{s.t.}~ |\textbf {$\phi$}^{\text{t}}_{\text{u},n}|^2 + |\textbf {$\phi$}^{\text{r}}_{\text{u},n}|^2 = 1, ~\forall n\in \mathcal{N},\\
&\hspace{-2.15cm} z_{\text{u},i} = \mathbf{\beta}_{ij} \times  z_{\text{u},j},~\forall i,j \in \mathcal{K}_\text{t} \cup \mathcal{K}_\text{r}, \\
&\ \hspace{-2.25cm} ||\textbf{V}_\text{u}||_F^2\leq 1.
\end{align}
\end{subequations}
Given \( \phi^{\mathcal{Y}}_{\text{u}} \), the matrix \(\Gamma^{\mathcal{Y}}_{\text{u}, k} \) is determined for each user, transforming the optimization problem into a sum of RQ with respect to \( \mathbf{v}_{\text{u}, k} \). Since the Rayleigh quotient is maximized when \( \mathbf{v}_{\text{u}, k} \) is aligned with \(\Gamma^{\mathcal{Y}}_{\text{u}, k} \), the optimal \( \mathbf{v}_{\text{u}, k} \) is the eigenvector corresponding to its largest eigenvalue \cite{zhang2013optimizing}.  To ensure constraint (\ref{phaseut}\rm{d}) is satisfied, the obtained beamforming vectors can be normalized without altering their alignment with the dominant eigen-structure. After deriving the active beamforming matrix, the phase shift vector \( \phi^{\mathcal{Y}}_{\text{u}} \) is updated by solving a structurally similar optimization problem to that of energy transfer in (\ref{etchannel}), with the only difference being the channel coefficients. The algorithm terminates once the objective function of \eqref{phaseut} no longer improves.

\paragraph{\text{TS}}
In the TS scenario, we can define an optimization problem similar to \eqref{phaseut} for transmission or reflection slot, where constraint (\ref{phaseut}\rm{b}) is modified as \( |\textbf{$\phi$}^{\mathcal{Y}}_{\text{u},n}| = 1 \) for all \( n \in \mathcal{N} \), \( \mathcal{Y} = \{\text{t}, \text{r}\} \). This revised problem can be solved using a similar approach.

\vspace{0.1cm}
\subsubsection{Joint Optimization of Downlink Transmission Phase Shift Vectors of STAR-RIS, and Beamforming Matrix; $\textbf{$\phi$}^{\mathcal{Y}}_{\text{d}}$ and $\textbf{V}_{\text{d}}$}
The consumed energy in downlink transmission can be rewritten from (\ref{AP_power}) and (\ref{AP_power1}), i.e.,
\begin{equation} \label{ECDown}
    P^{\text{ES}}_{\text{d}} \tau^{\text{ES}}_{\text{d}} = \frac{P^{\text{ES}}_{\text{d}}L_{\text{d}}}{\log_2\bigg{(}1 + {P^{\text{ES}}_{\text{d}} z_{\text{worst}}}\bigg{)}}.
\end{equation}
It is evident that maximizing the worst channel minimizes the energy consumption in downlink transmission. Depending on the STAR-RIS operation mode in downlink transmission, the following optimization problems are formulated to optimize both phase shift vectors of the STAR-RIS and beamforming matrix.

\paragraph{\text{ES}} 
Compared with the corresponding optimization problem for phase shift vectors and the beamforming matrix in energy transfer, we add a constraint to ensure that $z_{\text{worst}}$ represents the worst downlink channel.
\begin{subequations} {\label{phasedown}}
\begin{align}
\max_{\textbf{V}_{\text{d}},\phi^{\mathcal{Y}}_{\text{d}}}~~z_{\text{worst}},\\
&\ \hspace{-1.45cm} \text{s.t.}~ |\textbf {$\phi$}^{\text{t}}_{\text{d},n}|^2 + |\textbf {$\phi$}^{\text{r}}_{\text{d},n}|^2 = 1, ~\forall n\in \mathcal{N},\\
&\hspace{-0.8cm} z_{\text{d},i} = \mathbf{\beta}_{ij} \times  z_{\text{d},j},~\forall i,j \in \mathcal{K}_\text{t} \cup \mathcal{K}_\text{r}, \\
&\ \hspace{-0.99cm} ||\textbf{V}_\text{d}||_F^2\leq 1,\\
&\ \hspace{-0.9cm} z_{\text{worst}} \leq z_{\text{d},k}, \forall k \in \mathcal{K}_\text{t} \cup \mathcal{K}_\text{r}.
\end{align}
\end{subequations}
Then, the BCD iteratively updates $ \Phi_\text{d}^{\mathcal{Y}}$ and ${\textbf{V}}_{\text{d}, k}$ until the objective function of \eqref{phasedown} no longer improves.

{\bf{ Update $\phi^{\mathcal{Y}}_{\text{d}}$:}} Considering fixed beamforming matrix in the downlink transmission, the trace operator converts the above optimization problem into the following form.
\begin{subequations} {\label{dtchannel}}
\begin{flalign}
\max_{ \Phi_{\text{d}}^{\mathcal{Y}}}~~~ z_{\text{worst}}, &&\\
&\ \hspace{-1.9cm}\text{s.t.}~{\Phi}^{\mathcal{Y}}_{\text{d}} \succeq 0,~ &&\\
& \hspace{-1.3cm} \text{diag}({{\Phi}^{\text{t}}_{\text{d}}}) + \text{diag}({{\Phi}^{\text{r}}_{\text{d}}})=1^{\it{N}}, &&\\
& \hspace{-1.3cm} {\text{Rank}} ({\Phi}^{\mathcal{Y}}_{\text{d}}) =1,~ \mathcal{Y} = \{\text{t}, \text{r}\}, &&\\
& \hspace{-1.4cm} \frac{\sum_{ \mathcal{Y} = \{\text{t}, \text{r}\}} {\text{Tr}}\bigg{(}  \Phi_{\text{d}}^{\mathcal{Y}} \Lambda^{\mathcal{Y}}_{\text{d}, i} \bigg{)}}{\sum_{ \mathcal{Y} = \{\text{t}, \text{r}\}} {\text{Tr}}\bigg{(}  \Phi_{\text{d}}^{\mathcal{Y}} \Lambda^{\mathcal{Y}}_{\text{d}, j} \bigg{)}} = \beta_{ij}, ~\forall i,j \in \mathcal{K}_\text{t} \cup \mathcal{K}_\text{r}, &&\\
&\ \hspace{-1.4cm} z_{\text{worst}} \leq \sum_{ \mathcal{Y} = \{\text{t}, \text{r}\}} {\text{Tr}}\bigg{(} \Lambda^{\mathcal{Y}}_{\text{d}, k} \Phi_{\text{d}}^{\mathcal{Y}} \bigg{)},  ~\forall k \in \mathcal{K}_{\text{t}} \cup \mathcal{K}_{\text{r}}. &&
\end{flalign}
\end{subequations}
Excluding constraint (\ref{dtchannel}\rm{d}), the problem becomes a convex SDP optimization, which can be efficiently solved using CVX. Note that the Gaussian randomization is applied to obtain a rank-1 solution after the termination of the BCD algorithm.

{\bf{ Update $\textbf{V}_{\text{d}}$:}} Given $\phi^{\mathcal{Y}*}_{\text{d}}$ and considering trace properties, \eqref{phasedown} can be reduced into the following form.
\begin{subequations} {\label{dtchannel1}}
\begin{flalign}
\max_{{\textbf{V}}_{\text{d}, k}}~~~ z_{\text{worst}}, &&\\
&\ \hspace{-1.9cm}\text{s.t.}~{\textbf{V}}_{\text{d}, k} \succeq 0, ~\forall k \in \mathcal{K}_{\text{t}} \cup \mathcal{K}_{\text{r}}, &&\\
& \hspace{-1.3cm} \sum_{k = 1}^{K_{\text{t}} + K_{\text{r}}} {\text{Tr}} (\tilde {\textbf{V}}_{\text{d}, k}) \leq 1, &&\\
&\ \hspace{-1.3cm} {\text{Rank}} ({\textbf{V}}_{\text{d}, k}) =1, ~\forall k \in \mathcal{K}_{\text{t}} \cup \mathcal{K}_{\text{r}}, &&\\
&\ \hspace{-1.4cm} \frac{\sum_{ \mathcal{Y} = \{\text{t}, \text{r}\}} {\text{Tr}}\bigg{(} \Gamma^{\mathcal{Y}}_{\text{d}, i} {\textbf{V}}_{\text{d}, i} \bigg{)}}{\sum_{ \mathcal{Y} = \{\text{t}, \text{r}\}} {\text{Tr}}\bigg{(} \Gamma^{\mathcal{Y}}_{\text{d}, j} {\textbf{V}}_{\text{d}, j} \bigg{)}} = \beta_{ij}, \forall i,j \in \mathcal{K}_\text{t} \cup \mathcal{K}_\text{r}, &&\\
&\ \hspace{-1.4cm} z_{\text{worst}} \leq \sum_{ \mathcal{Y} = \{\text{t}, \text{r}\}} {\text{Tr}}\bigg{(} \Gamma^{\mathcal{Y}}_{\text{d}, k} \Phi_{\text{d}}^{\mathcal{Y}} \bigg{)},  \forall k \in \mathcal{K}_{\text{t}} \cup \mathcal{K}_{\text{r}}. &&
\end{flalign}
\end{subequations}
Similarly, CVX effectively solves this optimization problem, while the rank-1 constraint is addressed through Gaussian randomization after the last iteration of the BCD algorithm.

\paragraph{TS}
Since all elements of STAR-RIS operate in either transmission or reflection mode, the worst channel for each group must be optimized separately. However, due to user coupling in constraints (\ref{main_problem-ES_ES}\rm{i}) and (\ref{main_problem-ES_ES}\rm{j}), we heuristically optimize the phase shift vectors of the STAR-RIS along with the beamforming matrix on both sides through the following optimization problem:
\begin{subequations} {\label{phasedownTS}}
\begin{align}
\max_{\textbf{V}_{\text{d}},\phi^{\mathcal{Y}}_{\text{d}}}~~\sum_{\mathcal{Y} =\{\text{t}, \text{r}\}} z_{\text{worst}}^{\mathcal{Y}},\\
&\ \hspace{-2.6cm} \text{s.t.}~ |e^{j\theta^{\mathcal{Y}}_{\text{d},n}}|=1, ~\forall n \in \mathcal{N}, \mathcal{Y} =\{\text{t}, \text{r}\}, \\
&\hspace{-1.9cm}  z_{\text{d},i} = \mathbf{\beta}_{ij} \times  z_{\text{d},j},~\forall i,j \in \mathcal{K}_\text{t} \cup \mathcal{K}_\text{r}, \\
&\ \hspace{-2cm} ||\textbf{V}_\text{d}||_F^2\leq 1,\\
&\ \hspace{-2cm} z_{\text{worst}}^{\text{t}} \leq z_{\text{d},k}, \forall k \in \mathcal{K}_\text{t},\\
&\ \hspace{-2cm} z_{\text{worst}}^{\text{r}} \leq z_{\text{d},k}, \forall k \in \mathcal{K}_\text{r}.
\end{align}
\end{subequations}
Similar to \eqref{phasedown}, BCD can be utilized to iteratively optimize the STAR-RIS phase shift vectors and the beamforming matrix for downlink transmission by solving two SDP optimization problems.

\subsection{Joint Optimization of Time and Power for All Phases, Local Processing Time, Along with the Computation Frequency for Each User; $\tau_{\text{e}}, \tau_{{\text{l}},k}, \tau^{\mathcal{S}}_{{\text{u}},k}, \tau^{\mathcal{S}}_{\text{d}}, f_{k}, P_{\text{e}}, p_{\text{u}, k}$, and $P^{\mathcal{S}}_{\text{d}}$}

Depending on the STAR-RIS mode in either uplink or downlink transmission, we optimize time and power for all phases, local processing time, as well as the user's computation frequency for each scenario, while the phase shift vectors of STAR-RIS and the beamforming matrices are given.

\begin{algorithm} [b!]
\setlength{\abovecaptionskip}{0pt}
\setlength{\belowcaptionskip}{0pt}
\caption{ ES-ES }
\label{I}
\begin{algorithmic}[1]
     \State Initialize $\tau^{\text{ES}}_{\text{d}}$ with $P^{\text{ES}}_{\text{d}} = P_{\text{max}}$, $\tau_{{\text{l}},k}=\tau^{\text{ES}}_{{\text{u}},k}(0) = \frac{T - \tau^{\text{ES}}_{\text{d}}}{2}$ for all users.

     \While{$|T - \tau_{{\text{l}},k}  - \tau^{\text{ES}}_{{\text{u}},k}(\iota) - \tau^{\text{ES}}_{\text{d}}| \geq \epsilon, \forall k$,}
     \While{$|\tau^{\text{ES}}_{{\text{u}},k}(\iota)$ - $\tau^{\text{ES}}_{{\text{u}},k}(\iota - 1 )| \geq \epsilon, \forall k$,}
     
    \State Compute the corresponding \( p_{\text{u}, k}, \forall k \) by solving 
    
    the non-homogeneous linear equations in (\ref{R_u_ES}) and
    
    (\ref{main_problem-ES_ES}\rm{f}), subject to (\ref{main_problem-ES_ES}\rm{l}), using linear programming.

      \State Initialize  $\tau_{\text{e}} = \epsilon$.

     \While{ $\tau_{\text{e}} \leq T$,}

           ~~~~~~~~~~ Obtain the corresponding  $\tau_{{\text{l}},k}$ and $f_{k}$ $\forall k$ 
           
       ~~~~~~~~~~~from (\ref{main_problem-ES_ES}\rm{d}) and (\ref{main_problem-ES_ES}\rm{e}).

        \If {$\underset{k}{\min} (\tau_{{\text{l}},k}) \geq \tau_{\text{e}}$ \& $\tau_{{\text{l}},k} \leq T - \tau^{\text{ES}}_{{\text{u}},k} - \tau^{\text{ES}}_{\text{d}}, \forall k$,}

           ~~~~~~~~~~\textbf{break;}
           
           \Else

           ~~~~~~~~~~$\tau_{\text{e}} = \tau_{\text{e}} + \epsilon$,
           
           \EndIf           

     \EndWhile

     \State Update $\iota = \iota + 1$ and $\tau^{\text{ES}}_{{\text{u}},k}(\iota)$ from  (\ref{equalT}),
     
     \EndWhile

     \State Initialize $P^{\text{ES}}_{\text{d}} = \epsilon$.

     \While{ $P^{\text{ES}}_{\text{d}} \leq P_{\text{max}}$,}

      \If{$(\frac{L_{\text{d}}}{\log_2\left(1 + P^{\text{ES}}_{\text{d}} z_{\text{worst}}\right)}) \leq T - \tau_{{\text{l}},k} - \tau^{\text{ES}}_{{\text{u}},k}(\iota), \forall k$,}
    
      ~~~~~~\textbf{break;}


      \Else

       ~~~~~$P^{\text{ES}}_{\text{d}} = P^{\text{ES}}_{\text{d}} + \epsilon$,

     \EndIf

    \EndWhile
    \State Obtain the corresponding $\tau^{\text{ES}}_{\text{d}}$ from (\ref{SubtdPd}\rm{d}).

    \EndWhile

\end{algorithmic}
\end{algorithm}

\subsubsection{ES-ES}
Under given phase shift vectors of STAR-RIS and beamforming matrices for all phases, (\ref{main_problem-ES_TS}) can be expressed as:
\begin{subequations} \label{main_problem-ES_ES1}
\begin{align}
\min_{\tilde{\mathcal{V}}_1} \quad & P_{\text{max}} \tau_{\text{e}} + P^{\text{ES}}_{\text{d}} \tau^{\text{ES}}_{\text{d}}, \\&\ \hspace{-1.4cm} \text{s.t.}~ (\ref{main_problem-ES_ES}\rm{b} - \ref{main_problem-ES_ES}\rm{g}), (\ref{main_problem-ES_ES}\rm{l} - \ref{main_problem-ES_ES}\rm{m}).
\end{align}
\end{subequations}
where $\tilde{\mathcal{V}}_1 = \{ \tau_{\text{e}},\tau_{{\text{l}},k},\tau^{\text{ES}}_{{\text{u}},k}, \tau^{\text{ES}}_{\text{d}}, p_{\text{u}, k}, P^{\text{ES}}_{\text{d}}, f_k\}$.

In the energy transfer phase, the lower bound of harvesting time represents the minimum consumed energy. Consequently, equality must hold in (\ref{main_problem-ES_ES}\rm{d}) for all users, ensuring  
\begin{equation} \label{t_{e,k}}
   \eta P_{\text{e}} z_{\text{e},k} \tau_{\text{e}} - p_{\text{u}, k} \tau^{\text{ES}}_{{\text{u}},k} - a f^3_{k} \tau_{{\text{l}},k} = 0, \quad \forall k \in \mathcal{K}_\text{t} \cup \mathcal{K}_\text{r}.
\end{equation}

Reducing energy consumption for each user during uplink transmission and local processing directly decreases the required harvested energy for that user. From  \eqref{local}, (\ref{main_problem-ES_ES}\rm{e}), and \eqref{equivalent}, minimizing energy consumption during these phases necessitates maximizing $\tau_{\text{l}, k}$ and $\tau^{\text{ES}}_{\text{u}, k}$, respectively.  

For the third phase, maximizing downlink transmission time minimizes energy consumption, as indicated by \eqref{ECDown} and (\ref{main_problem-ES_ES}\rm{g}).  

Considering all these, prolonging the duration of local processing for each user, uplink transmission for each user, and downlink transmission effectively reduces total energy consumption. This implies that equality should hold in (\ref{main_problem-ES_ES}\rm{b}) for all users, i.e.,  
\begin{equation} \label{equalT}
    T - \tau_{{\text{l}},k} - \tau^{\text{ES}}_{{\text{u}},k} - \tau^{\text{ES}}_{\text{d}} = 0, \quad \forall k \in \mathcal{K}_\text{t} \cup \mathcal{K}_\text{r}.
\end{equation}

Algorithm \ref{I} can determine the optimized point where we allocate the resources for energy harvesting, local processing, and uplink transmission using the ODSA over $\tau_{\text{e}}$ for an initialized $\tau^{\text{ES}}_{\text{d}}$. Subsequently, the downlink transmission time and power are updated by solving the following optimization problem using another ODSA over $P^{\text{ES}}_{\text{d}}$. Note that the entire algorithm iterates until \(T - \tau_{{\text{l}},k}  - \tau^{\text{ES}}_{{\text{u}},k} - \tau^{\text{ES}}_{\text{d}} \leq \epsilon, \forall k\) is satisfied.
\begin{subequations} \label{SubtdPd}
\begin{align}
\min_{P^{\text{ES}}_{\text{d}}, \tau^{\text{ES}}_{\text{d}}} \quad & P^{\text{ES}}_{\text{d}} \tau^{\text{ES}}_{\text{d}}, \\ 
\text{s.t.} \quad & \tau^{\text{ES}}_{\text{d}} \leq T - \tau^{*}_{{\text{l}},k} - \tau^{\text{ES*}}_{{\text{u}},k}, \quad \forall k \in \mathcal{K}_{\text{t}} \cup \mathcal{K}_{\text{r}}, \\
& R^{\text{ES}}_{\text{d}} = \log_2\bigg(1 + {P^{\text{ES}}_{\text{d}} z_{\text{worst}}}\bigg), \\
& L_{\text{d}} = R^{\text{ES}}_{\text{d}} \tau^{\text{ES}}_{\text{d}}, \\
& P^{\text{ES}}_{\text{d}} \leq P_{\text{max}}.
\end{align}
\end{subequations}

 \begin{algorithm} [b!]
\setlength{\abovecaptionskip}{0pt}
\setlength{\belowcaptionskip}{0pt}
\caption{ TS Downlink }
\label{II}
\begin{algorithmic}[1]
     \State Initialize $P^{\text{t}}_{\text{d}} = \epsilon$ and Interval = $P_{\text{max}}$.
     
     \While{$P^{\text{t}}_{\text{d}} \leq P_{\text{max}}$}
     \vspace{0.1cm}
      
      $P^{\text{r}}_{\text{d}}(\text{low}) = \epsilon$, $P^{\text{r}}_{\text{d}}(\text{up}) = P_{\text{max}} - P^{\text{t}}_{\text{d}} - \epsilon$, and 
      
      $P^{\text{r}}_{\text{d}}(\text{mid}) =  \frac{P^{\text{r}}_{\text{d}}(\text{low}) +  P^{\text{r}}_{\text{d}}(\text{up})}{2}$.

       \vspace{0.1cm}
      \While{$\text{Interval} \geq \epsilon$}

      $\text{Mean value} = \frac{\frac{P^{\text{r}}_{\text{d}}(\text{up})} {\log_2\left(1 + P^{\text{r}}_{\text{d}}(\text{up}) z^{\text{r}}_{\text{worst}}\right)} - \frac{P^{\text{r}}_{\text{d}}(\text{mid})} {\log_2\left(1 + P^{\text{r}}_{\text{d}}(\text{mid}) z^{\text{r}}_{\text{worst}}\right)}}{P^{\text{r}}_{\text{d}}(\text{up}) - P^{\text{r}}_{\text{d}}(\text{mid})}$.

      \If {$\text{Mean value} \geq 0$}

      ~~~~~$P^{\text{r}}_{\text{d}}(\text{up}) = P^{\text{r}}_{\text{d}}(\text{mid})$, $P^{\text{r}}_{\text{d}}(\text{mid}) = \frac{P^{\text{r}}_{\text{d}}(\text{low}) +  P^{\text{r}}_{\text{d}}(\text{up})}{2}$, 

      \Else

       ~~~~~$P^{\text{r}}_{\text{d}}(\text{low}) = P^{\text{r}}_{\text{d}}(\text{mid})$, $P^{\text{r}}_{\text{d}}(\text{mid}) = \frac{P^{\text{r}}_{\text{d}}(\text{low}) +  P^{\text{r}}_{\text{d}}(\text{up})}{2}$, 

      \EndIf

      $\text{Interval} = P^{\text{r}}_{\text{d}}(\text{up}) - P^{\text{r}}_{\text{d}}(\text{low})$.

      \EndWhile

      \If{$\frac{L_{\text{d}}}{\log_2\left(1 + P^{\text{t}}_{\text{d}} z^{\text{t}}_{\text{worst}}\right)} + \frac{L_{\text{d}}}{\log_2\left(1 + P^{\text{r}}_{\text{d}}(\text{mid}) z^{\text{r}}_{\text{worst}}\right)} \leq T - \tau^{*}_{{\text{l}},k} - \tau^{\text{ES*}}_{{\text{u}},k}$,}
      \vspace{0.1em}
    
      ~~ \textbf{break;}

      \Else

      ~~$P^{\text{t}}_{\text{d}} = P^{\text{t}}_{\text{d}} + \epsilon$.

     \EndIf

     \EndWhile

    \State Obtain the corresponding $\tau^{\text{t}}_{\text{d}}$ and  $\tau^{\text{r}}_{\text{d}}$ from (\ref{SubtdPd1}\rm{c}) and (\ref{SubtdPd1}\rm{d}).
    
\end{algorithmic}
\end{algorithm}
By substituting (\ref{SubtdPd}\rm{c}) and (\ref{SubtdPd}\rm{d}) into the objective function, the optimization problem can be simplified as follows:
\begin{subequations} \label{Ding1}
\begin{align}
\min_{P^{\text{ES}}_{\text{d}}} \quad & \frac{P^{\text{ES}}_{\text{d}} L_{\text{d}}} {\log_2\left(1 + P^{\text{ES}}_{\text{d}} z_{\text{worst}}\right)}, \\ 
\text{s.t.} \quad & \frac{L_{\text{d}}}{\log_2\left(1 + P^{\text{ES}}_{\text{d}} z_{\text{worst}}\right)} \leq T - \tau^{*}_{{\text{l}},k} - \tau^{\text{ES*}}_{{\text{u}},k}, \forall k \in \mathcal{K}_\text{t} \cup \mathcal{K}_\text{r},\\
& P^{\text{ES}}_{\text{d}} \leq P_{\text{max}}.
\end{align}
\end{subequations}
Since the objective function is strictly increasing with respect to \(P^{\text{ES}}_{\text{d}}\), an ODSA over \(P_{\text{max}}\) can efficiently determine the optimal downlink transmission power. Indeed, the smallest value of \(P^{\text{ES}}_{\text{d}}\) that satisfies the constraint in (\ref{Ding1}\rm{b}) will be the optimized downlink transmission power. The corresponding time slot \(\tau^{\text{ES}}_{\text{d}}\) is then computed using (\ref{SubtdPd}\rm{c}) and (\ref{SubtdPd}\rm{d}).

\begin{theorem} \label{th1em}
    The optimal point is unique, if feasible. 
\end{theorem}

\begin{IEEEproof}
The uniqueness of the transmission time and power in downlink transmission is evident from (\ref{AP_power}) and (\ref{AP_power1}). Since (\ref{R_u_ES}) is a non-homogeneous system for all users, the uniqueness of \( p_{\text{u}, k} \) for all \( k \) is guaranteed. Furthermore, \( \tau^{\text{ES}}_{\text{u}, k} \) is uniquely determined by (\ref{t_k1}). Following this, the uniqueness of \( \tau_{\text{l}, k} \) and \( f_k \) for all \( k \) is ensured by \eqref{equalT} and (\ref{main_problem-ES_ES}\rm{e}). Therefore, the uniqueness of \( \tau_{\text{e}} \) is also guaranteed from \eqref{t_{e,k}}. It should be mentioned that (\ref{main_problem-ES_ES}\rm{c}) and (\ref{main_problem-ES_ES}\rm{l}) clarify the feasibility of the optimal point.
 
\end{IEEEproof}

\subsubsection{ES-TS}
Under given phase shift vectors of STAR-RIS and beamforming matrices for all phases, (\ref{main_problem-ES_TS}) can be reduced to
\begin{subequations} \label{main_problem-ES_TS1}
\begin{align}
\min_{\tilde{\mathcal{V}}_2} \quad & P_{\text{max}} \tau_{\text{e}} + \sum_{\mathcal{Y} =\{\text{t}, \text{r}\}} P^{\mathcal{Y}}_{\text{d}} \tau^{\mathcal{Y}}_{\text{d}}, \\&\ \hspace{-1.4cm} \text{s.t.}~ (\ref{main_problem-ES_ES}\rm{c} - \ref{main_problem-ES_ES}\rm{f}), (\ref{main_problem-ES_ES}\rm{l}), (\ref{main_problem-ES_TS}\rm{b} - \ref{main_problem-ES_TS}\rm{c}), (\ref{main_problem-ES_TS}\rm{f}),
\end{align}
\end{subequations}
where $\tilde{\mathcal{V}}_2 = \{ \tau_{\text{e}},\tau_{{\text{l}},k},\tau^{\text{ES}}_{{\text{u}},k}, \tau^{\mathcal{Y}}_{\text{d}}, p_{\text{u}, k}, P^{\mathcal{Y}}_{\text{d}}, f_k$\}. 

In this scenario, the downlink transmission phase differs from the ES-ES case. Consequently, the only modification in Algorithm \ref{I} involves updating the downlink transmission time and power over the following optimization problem.
\begin{subequations} \label{SubtdPd1}
\begin{align}
\min_{P^{\text{t}}_{\text{d}}, P^{\text{r}}_{\text{d}}, \tau^{\text{t}}_{\text{d}},  \tau^{\text{r}}_{\text{d}}} \quad & \sum_{\mathcal{Y} =\{\text{t}, \text{r}\}} P^{\mathcal{Y}}_{\text{d}} \tau^{\mathcal{Y}}_{\text{d}}, \\&\ \hspace{-1.3cm} \text{s.t.}~  \sum_{\mathcal{Y} =\{\text{t}, \text{r}\}} \tau^{\mathcal{Y}}_{\text{d}} \leq T - \tau^{*}_{{\text{l}},k} - \tau^{\text{ES*}}_{{\text{u}},k},~\forall k \in \mathcal{K}_{\text{t}} \cup \mathcal{K}_{\text{r}},\\
& \hspace{-0.6cm} R^{\text{TS}, \mathcal{Y}}_{\text{d}} = \log_2\bigg{(}1 + {P^{\mathcal{Y}}_{\text{d}} z^{\mathcal{Y}}_{\text{worst}}}\bigg{)}, ~ \mathcal{Y} = \{\text{t}, \text{r}\},  \\
&\hspace{-0.6cm} L_{\text{d}} = R^{\text{TS}, \mathcal{Y}}_{\text{d}} \tau^{\mathcal{Y}}_{\text{d}}, ~ \mathcal{Y} = \{\text{t}, \text{r}\},\\
&\hspace{-0.7cm}  \sum_{\mathcal{Y} =\{\text{t}, \text{r}\}} P^{\mathcal{Y}}_{\text{d}} \leq P_{\text{max}}.
\end{align}
\end{subequations}
By substituting (\ref{SubtdPd1}\rm{c}) and (\ref{SubtdPd1}\rm{d}) into the above optimization problem, it can be reformulated as
\begin{subequations} \label{Ding11}
\begin{align}
\min_{P^{\text{t}}_{\text{d}}, P^{\text{r}}_{\text{d}}} \quad & \sum_{\mathcal{Y} = \{\text{t}, \text{r}\}}  \frac{P^{\mathcal{Y}}_{\text{d}}} {\log_2\left(1 + P^{\mathcal{Y}}_{\text{d}} z^{\mathcal{Y}}_{\text{worst}}\right)}, \\&\ \hspace{-0.65cm} \text{s.t.}~  \sum_{\mathcal{Y} = \{\text{t}, \text{r}\}} \frac{L_{\text{d}}}{\log_2\left(1 + P^{\mathcal{Y}}_{\text{d}} z^{\mathcal{Y}}_{\text{worst}}\right)} \leq T - \tau^{*}_{{\text{l}},k} - \tau^{\text{ES*}}_{{\text{u}},k}, \forall k,\\
& \sum_{\mathcal{Y} =\{\text{t}, \text{r}\}} P^{\mathcal{Y}}_{\text{d}} \leq P_{\text{max}}.
\end{align}
\end{subequations}
This optimization problem can be efficiently solved by integrating the BA and ODSA, as outlined in Algorithm \ref{II}, assuming $z^{\text{t}}_{\text{worst}} \geq z^{\text{r}}_{\text{worst}}$.

\subsubsection{TS-ES}
Given the phase shift vectors and beamforming matrices for all phases, the TS-ES scenario can be expressed as follows.
\begin{subequations} \label{main_problem-TS_ES1}
\begin{align}
\min_{\tilde{\mathcal{V}}_3} \quad & P_{\text{max}} \tau_{\text{e}} + P^{\text{ES}}_{\text{d}} \tau^{\text{ES}}_{\text{d}}, \\&\ \hspace{-1.4cm} \text{s.t.}~ (\ref{main_problem-ES_ES}\rm{c} - \ref{main_problem-ES_ES}\rm{g}), (\ref{main_problem-ES_ES}\rm{l} - \ref{main_problem-ES_ES}\rm{m}), (\ref{main_problem-TS_ES}\rm{b}),
\end{align}
\end{subequations}
where $\tilde{\mathcal{V}}_3 = \{ \tau_{\text{e}},\tau_{{\text{l}},k},\tau^{\text{TS}}_{{\text{u}},k}, \tau^{\text{ES}}_{\text{d}}, p_{\text{u}, k}, P^{\text{ES}}_{\text{d}}, f_k\}$.

In this scenario, the key difference from problem \eqref{main_problem-ES_ES1} lies in the system of linear equations in (\ref{main_problem-TS_ES}\rm{b}). Therefore, holding equality in (\ref{main_problem-TS_ES}\rm{b}) while maintaining the maximization of the uplink transmission time for each user is satisfied if and only if \( \tau_{{\text{l}},k} =  \tau_{{\text{l}}}, \forall k \) and \( \tau^{\text{TS}}_{{\text{u}},k} =  \tau^{\text{TS}}_{{\text{u}}}, \forall k \). Given this condition, the proposed approach for the ES-ES scenario, i.e., Algorithm \ref{I}, remains applicable for determining the optimal solution, with modifications in overall delay constraint throughout the entire algorithm, i.e., $\tau^{\text{TS}}_{\text{u}} = \frac {T - \tau_{{\text{l}}} -\tau^{\text{ES}}_{\text{d}}}{2}$.
 
\vspace{0.25 cm} 
\subsubsection{TS-TS}
Driven from (\ref{main_problem-TS_TS}) under given phase shift vectors of STAR-RIS and beamforming matrices for each phase, the following optimization problem is formulated.
\begin{subequations} \label{main_problem-TS_TS1}
\begin{align}
\min_{\tilde{\mathcal{V}}_4} \quad & P_{\text{max}} \tau_{\text{e}} + \sum_{\mathcal{Y} =\{\text{t}, \text{r}\}} P^{\mathcal{Y}}_{\text{d}} \tau^{\mathcal{Y}}_{\text{d}}, 
\\&\ \hspace{-1.7cm}
\begin{aligned}
& \substack{\text{\normalsize (\ref{main_problem-ES_ES}\rm{c}),(\ref{main_problem-ES_ES}\rm{e}), (\ref{main_problem-ES_ES}\rm{l}), (\ref{main_problem-ES_TS}\rm{c}),} \\[3pt] \text{\normalsize (\ref{main_problem-ES_TS}\rm{f}), (\ref{main_problem-TS_ES}\rm{c} - \ref{main_problem-TS_ES}\rm{d}), (\ref{main_problem-TS_TS}\rm{b}),}}
\end{aligned} 
\end{align}
\end{subequations}
where $\tilde{\mathcal{V}}_4 = \{ \tau_{\text{e}},\tau_{{\text{l}},k},\tau^{\text{TS}}_{{\text{u}},k}, \tau^{\mathcal{Y}}_{\text{d}}, p_{\text{u}, k}, P^{\mathcal{Y}}_{\text{d}}, f_k\}$. 

In this scenario, we modify Algorithm \ref{I} in updating the downlink transmission time and power using Algorithm \ref{II} while incorporating modifications to the overall delay constraint throughout the entire algorithm, i.e., $\tau^{\text{TS}}_{\text{u}} = \frac{T - \tau_{{\text{l}}} - \tau^{\text{t}}_{\text{d}} - \tau^{\text{r}}_{\text{d}}}{2}$.

\subsection{Convergence and Computational Complexity Analysis}
\subsubsection{Convergence}
The convergence of optimizing phase shift vectors of STAR-RIS and beamforming matrices in each phase for all scenarios is guaranteed due to the termination of the BCD algorithm across two consecutive convex optimizations once the corresponding objective function no longer improves. Moreover, the uniqueness of the optimized solution for Block B, as established in Theorem \ref{th1em}, together with the employment of ODSA/BA, guarantees the convergence of Algorithm \ref{I}. Note that the overall convergence is clear since the blocks are independent.

\begin{figure}[htbp]
    \centering
    \includegraphics[width=\textwidth,width=\linewidth]{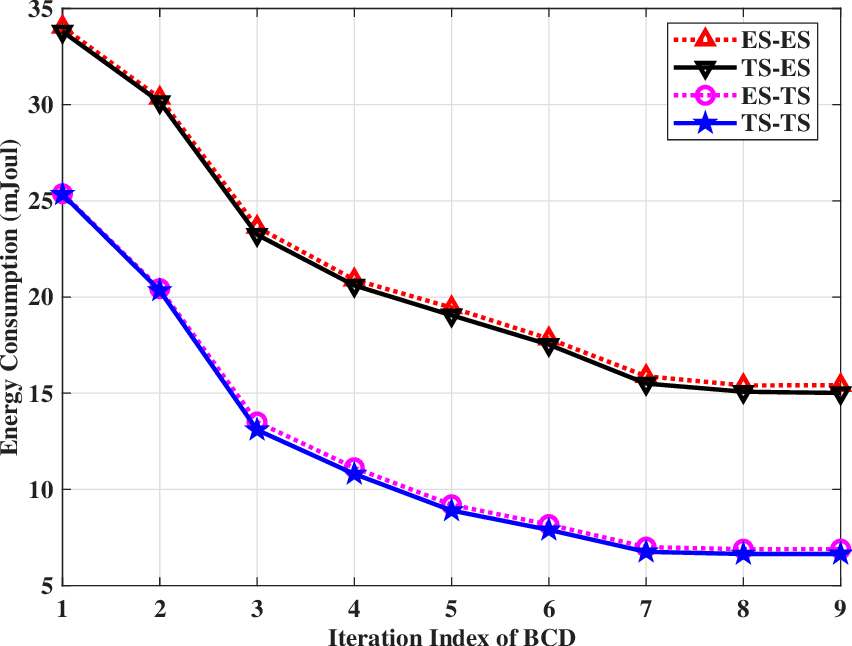}
    \caption{{Energy consumption vs. iteration index for $K = 4$, $N = 60$, $M = 4$, $\eta = 0.8$, $P_{\text{max}}= 10 $ W, and $T = 10$ s.}}
    \label{fig:4con}
  \end{figure}

To illustrate practical convergence, Fig.~\ref{fig:4con} plots the energy consumption versus iteration index, showing that the proposed BCD-based algorithm converges rapidly within a few iterations.

\subsubsection{Computational Complexity}
Regardless of STAR-RIS operation mode, the computation complexity of optimization for phase shift vectors of STAR-RIS and beamforming matrix in downlink transmission is the worst case in Block A and is formulated as $ \mathcal{O} (4N^4 (2N^2+N+K_{\text{t}}+K_{\text{r}}+3) \sqrt(N+K_{\text{t}}+K_{\text{r}}+3) \log_2 (\frac{1}{\epsilon})) + \mathcal{O} (M^4 (M^2+K_{\text{t}}+K_{\text{r}}+4) \sqrt(K_{\text{t}}+K_{\text{r}}+4) \log_2 (\frac{1}{\epsilon}))$ \cite{10032501}.

Algorithm~\ref{I} performs two ODSAs for the ES-ES and TS-ES scenarios, resulting in a complexity of \( \mathcal{O}\left(\frac{T}{\epsilon} +\frac{P_{\text{max}}}{\epsilon}\right) \). In the ES-TS and TS-TS scenarios, Algorithm~\ref{I} is modified in updating the downlink transmission time and power in each iteration using Algorithm~\ref{II}, which incorporates BA on top of ODSA. The overall complexity of the modified algorithm is then given by \( \mathcal{O}\left(\frac{T}{\epsilon} + \frac{P_{\text{max}}}{2\epsilon} \times \log (\frac{P_{\text{max}}}{2\epsilon})\right) \) \cite{6786060, 10400811}.

\section{Simulation Results}
Although our proposed algorithm performs effectively for different numbers of users in each group, we simplify the simulation by setting \( K_{\text{t}} = K_{\text{r}} = K = 4 \). The STAR-RIS is positioned at the origin, whereas the AP is located at coordinates \( (-20\, \text{m}, 20\, \text{m}) \). Group A users are distributed along the boundary of the top right quadrant of the circle with a radius of \( 20 \, \text{m} \) centered at the STAR-RIS, while Group B users are distributed along the boundary of the bottom left quadrant of that circle. All channel parameters are composed of both small-scale and large-scale fading. The path loss is modeled as $L = {L_0 d^{-\rho}}$, where \( \rho \) is the path loss exponent, \( L_0 = -30\), and \( d \) represents the distance from STAR-RIS either to the AP or users. The communication channels in each phase between user $k$ and STAR-RIS are given by $\Delta_k = L \left( \sqrt{\frac{\varrho}{\varrho + 1}} \Delta_{k}^{\text{LoS}} + \sqrt{\frac{1}{\varrho + 1}} \Delta_{k}^{\text{NLoS}} \right)$, while that between the STAR-RIS and the AP is $\Delta = L \left( \sqrt{\frac{\varrho}{\varrho + 1}} \Delta^{\text{LoS}} + \sqrt{\frac{1}{\varrho + 1}} \Delta^{\text{NLoS}} \right)$. The terms \( \varrho \), \( \Delta^{\text{LoS}} \), and \( \Delta^{\text{NLoS}} \) correspond to the Rician factor, line of sight (LoS), and non-line of sight (NLoS) components of the channel, respectively. For comparison, a reflecting-only RIS and a transmitting-only RIS, each with \( \frac{N}{2} \) elements, are placed at the STAR-RIS location to achieve full-space coverage, represented as ``Conventional RIS''. Indeed, the baseline scenario involves utilizing MS STAR-RIS across all phases, with a predetermined number of STAR-RIS elements allocated for both transmission and reflection regions. All simulation results are averaged over \( 10^3 \) random trials with an error tolerance of $\epsilon = 10^{-5}$. Table \ref{tab:3} summarizes the main simulation parameters.
\begin{table}[htbp]
    \centering
    \resizebox{9cm}{!}{ 
    \begin{tabular}{p{5.6cm} p{2.45cm}}
    \hline
     Parameters & Values \\
    \hline
     Maximum power of AP, $P_{\text{max}}$ & $10$ W \\
     Maximum power for user $k$, $p^{\text{max}}_{k}$ & $0.1$ W \\
     Overall delay $T$ & $10$ s \\
     Bandwidth &   $2$ MHz  \\
     Number of users for each group, $K$   & $4$   \\
     Number of reflection elements, $N$ &   $60$   \\
     Number of the AP antenna, $M$ &   $4$   \\
     Path loss exponents,  $\rho$ & $3$/$3.5$ \\
     Reference distance and path loss, $d_0$/$L_0$ & $1$ m/$30$ dB \\
     Computational complexity, $C_k$ &   $300$ cycle/bit  \\
     Rician factor, $\varrho$ & $5$  \\
     Energy conversion efficiency, $\eta$ & $0.8$  \\
     Noise power, $\sigma_0$ & $-174$ dBm/Hz \\
     Energy coefficient, $a$ & $10^{-28}$ \\
     Processed data size, $L_{{\text{l}},k}$ & [$0.1$ - $1$]~Mbits \\
     Uploaded data size, $L_{{\text{u}},k}$ & [$1$ - $10$]~Kbits \\
     Downloaded data size, $L_{\text{d}}$ & $1$ Mbits \\
    \hline
    \end{tabular}
    }
    \caption{{Default Simulation Parameters \cite{ 10032501, 10032506, 10028982}}}
    \label{tab:3}
\end{table}

 \begin{figure}[b!]
    \centering
    \includegraphics[width=\textwidth,width=\linewidth]{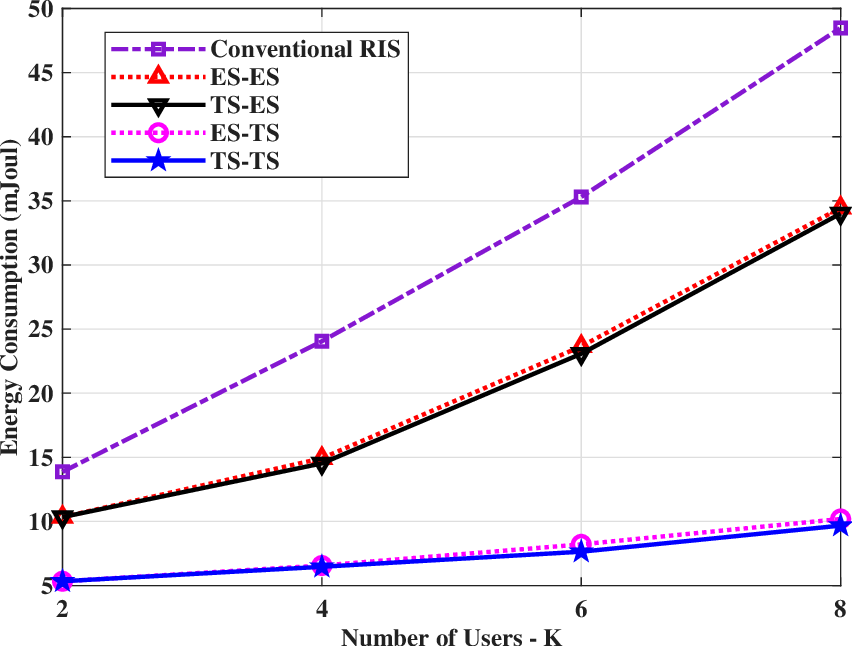}
    \caption{Energy consumption vs.  Number of users in each group for $N = 60$, $M = 4$, $\eta = 0.8$, $P_{\text{max}}= 10 $ W, and $T = 10$ s.}
    \label{fig:3k}
  \end{figure}
  
Figure \ref{fig:3k} compares energy consumption across various scenarios, considering different numbers of users in each group. As anticipated, the energy consumption for all schemes increases with the number of users due to heightened intra-group and inter-group interference. Scenarios with TS STAR-RIS in uplink transmission demonstrate superior performance for two key reasons: first, TS STAR-RIS in uplink transmission, each user experiences only intra-group interference, whereas ES STAR-RIS in uplink requires more energy harvesting as it suffers from both intra-group and inter-group interference, resulting in higher overall energy consumption; second, TS STAR-RIS serves each group separately, allowing more flexible adaptation of STAR-RIS phase shift vectors for each group users during uplink transmission, while ES STAR-RIS shares the same phase shift vectors among all users in both groups, resulting in lower channel gains. The challenge of inter-interference and adapting STAR-RIS phase shift vectors in ES STAR-RIS for uplink becomes more pronounced with more users, as evidenced by the increasing performance gap between TS and ES. On the other hand, TS in downlink transmission proves more effective than ES due to its ability to flexibly adapt phase shift vectors for each group. The gap between scenarios with different STAR-RIS modes in downlink transmission is negligible, as the global model, the time, and energy spent on downlink transmission are minimal per communication round compared to those required for local processing and uplink transmission. Overall, TS-TS is the most energy-saving scheme, followed by TS-ES, ES-TS, ES-ES, and the Conventional RIS. The conventional RIS performs the worst as it operates like MS STAR-RIS, where half of the elements serve in transmission mode and the other half in reflection mode. This effectively mimics ES but with only half the elements, resulting in a weaker channel and higher energy consumption.

\begin{figure}[b!]
    \centering
    \includegraphics[width=\textwidth,width=\linewidth]{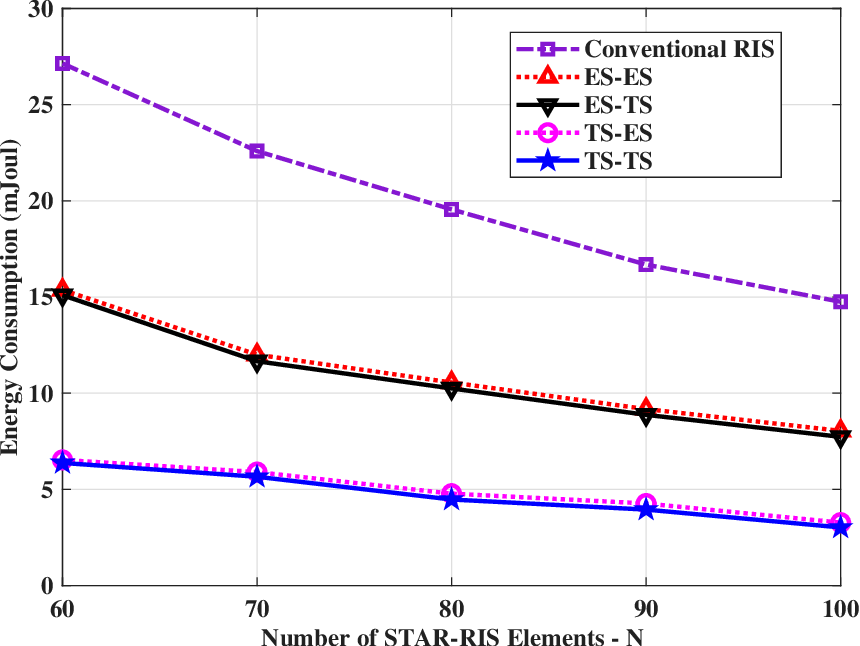}
    \caption{Energy consumption vs.  Number of STAR-RIS elements for $K = 4$, $M = 4$, $\eta = 0.8$, $P_{\text{max}}= 10 $ W, and $T = 10$ s.}
    \label{fig:4}
  \end{figure}

\begin{figure}[htbp] 
    \centering
    \includegraphics[width=\textwidth,width=\linewidth]{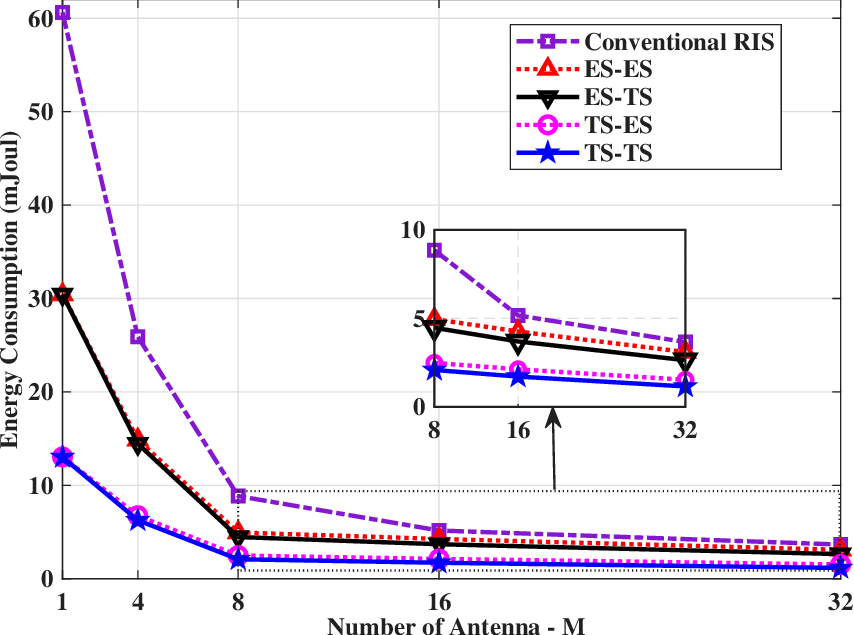}
    \caption{Energy consumption vs.  Number of AP antenna for $K = 4$, $N = 60$, $\eta = 0.8$, $P_{\text{max}}= 10 $ W, and $T = 10$ s.}
    \label{fig:5}
  \end{figure}

Figures \ref{fig:4} and \ref{fig:5} illustrate the energy consumption versus different numbers of STAR-RIS elements and AP antennas, respectively. A larger number of STAR-RIS elements or AP antennae enhances energy harvesting and improves both uplink and downlink data transmission, ultimately leading to greater energy savings for the system. This aligns with the observed decreasing trend across all scenarios. Notably, the TS-TS scheme achieves the lowest energy consumption, outperforming the other schemes.

\begin{figure}[htbp] 
    \centering
    \includegraphics[width=\textwidth,width=\linewidth]{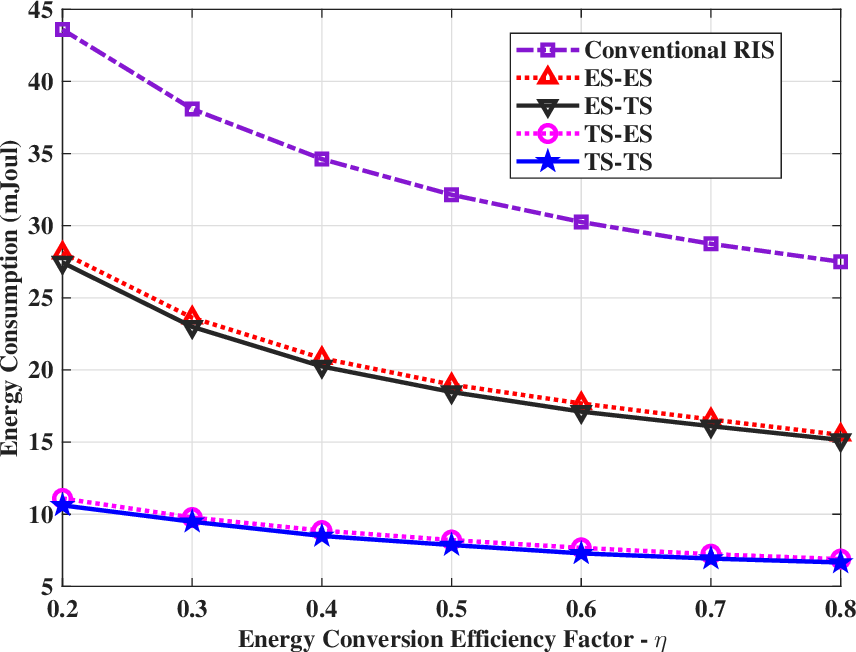}
    \caption{Energy consumption vs. Energy conversion efficiency factor for $K = 4$, $N = 60$, $M = 4$, $\eta = 0.8$, $P_{\text{max}}= 10 $ W, and $T = 10$ s.}
    \label{fig:6}
  \end{figure}

\begin{figure}[t!]
    \centering
    \includegraphics[width=\textwidth,width=\linewidth]{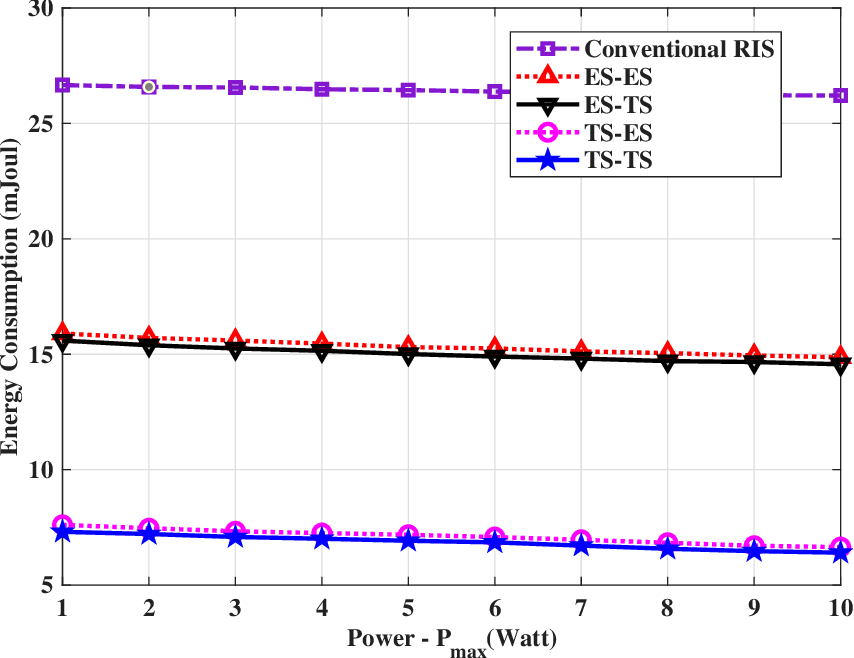}
    \caption{Energy consumption vs. Power of AP for $K = 4$, $N = 60$, $M = 4$, $\eta = 0.8$, and $T = 10$ s.}
    \label{fig:7}
  \end{figure}

Figures \ref{fig:6} and \ref{fig:7} reveal the energy consumption for varying energy conversion efficiency factors and AP power levels, respectively.  As expected, the performance of all schemes demonstrates a downward trend with the energy conversion efficiency factor and AP transmit power, owing to more harvested energy by each user, aligning with the findings of \cref{theorem1}. Increasing the energy conversion efficiency factor has a multiplicative effect on energy harvesting, leading to exponential-like gains in energy efficiency. In contrast, increasing the maximum power of the AP provides only a linear decrease in energy consumption since the energy harvesting time is reduced to minimize energy consumption. Consequently, the decrease in energy consumption due to the energy conversion efficiency factor is significantly faster than that achieved by increasing the power of AP. Once again, the TS-TS scheme performs the best, followed by TS-ES, ES-TS, ES-ES, and Conventional RIS.

 \begin{figure}[htbp]
    \centering
    \includegraphics[width=\textwidth,width=\linewidth]{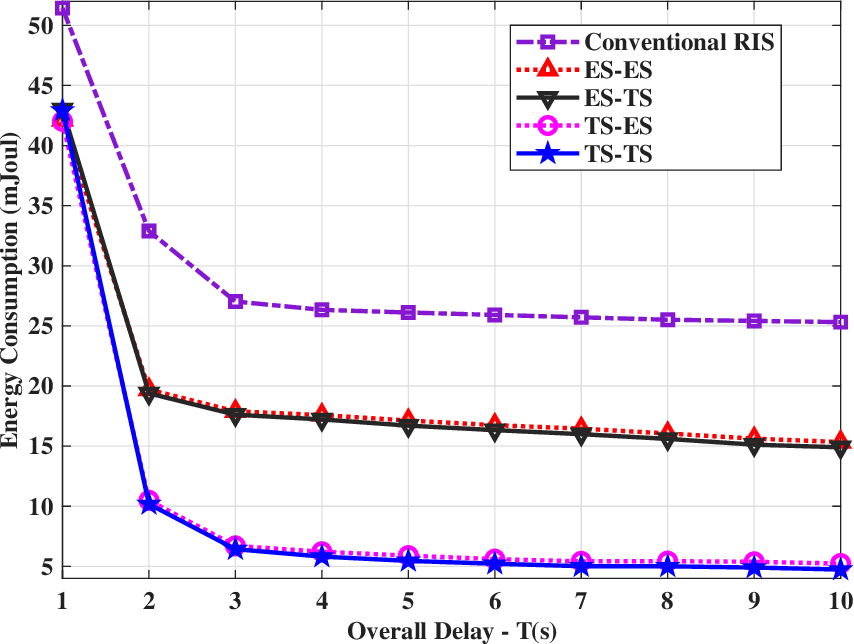}
    \caption{Energy consumption vs. Overall latency  for $K = 4$, $N = 60$, $M = 4$, $\eta = 0.8$, and $P_{\text{max}}= 10 $ W.}
    \label{fig:8}
  \end{figure}
Figure \ref{fig:8} illustrates the energy consumption under varying overall delays, with a downward trend across all scenarios. This is because the system gains more time for local processing, transmission, and interference management. This supports our reasoning that maximizing time for local processing, uplink, and downlink transmission reduces energy consumption. The ES STAR-RIS in the uplink benefits from time-sharing between both groups, allowing more time for uplink transmission. This advantage helps mitigate the negative impact of inter-interference in ES STAR-RIS uplink transmission, particularly at lower overall delays, where the gaps between scenarios with  ES STAR-RIS in uplink transmission are narrowed compared to scenarios with TS STAR-RIS in uplink transmission. However, with increasing delay, the gap widens as the inter-interference impact outweighs the benefits of time-sharing in ES STAR-RIS uplink transmission. The TS-TS scheme remains the most energy-saving scenario among all across various overall delays.

\begin{figure}[htbp]
    \centering
    \includegraphics[width=\textwidth,width=\linewidth]{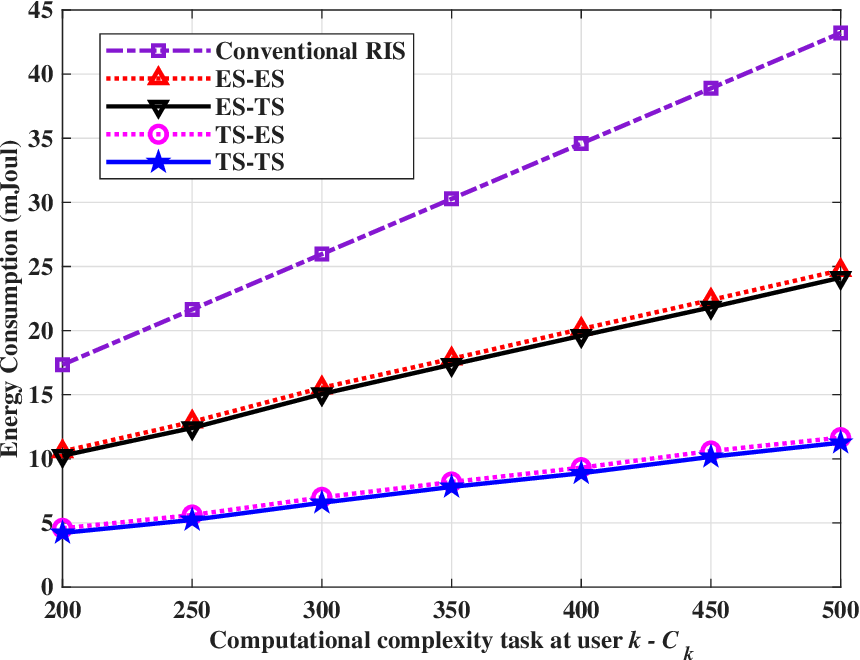}
    \caption{Energy consumption vs.  Computational complexity task at user $k$ for $K = 4$, $N = 60$, $M = 4$, $\eta = 0.8$, $P_{\text{max}}= 10 $ W, and $T = 10$ s.}
    \label{fig:9}
  \end{figure}

Figure \ref{fig:9} plots the energy consumption for different computational complexity tasks at user $k$ across all scenarios, exhibiting an upward trend as higher computational complexity tasks demand more energy for local processing. This exacerbates the challenge of handling inter-interference in scenarios with ES STAR-RIS in uplink transmission compared to those with TS STAR-RIS, leading to a widening gap between the two as computational complexity increases. The TS-TS scheme is still the best among all for different computational complexity tasks at user $k$.

\begin{figure}[htbp]
    \centering
    \includegraphics[width=\textwidth,width=\linewidth]{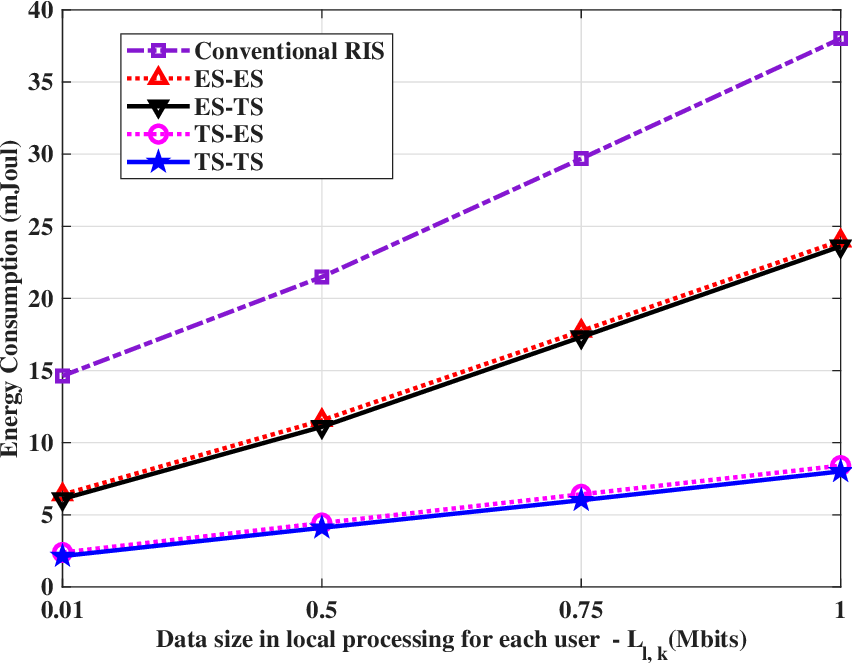}
    \caption{Energy consumption vs.  Data size in local processing for each user for $K = 4$, $N = 60$, $M = 4$, $\eta = 0.8$, $P_{\text{max}}= 10 $ W, and $T = 10$ s.}
    \label{fig:10}
  \end{figure}

Figure \ref{fig:10} presents how the energy consumption varies with data size in local processing for each user across all scenarios. A larger local processing data size for each user results in a bigger data size in uplink transmission and the global model, consequently requiring more energy consumption in all phases, as evidenced by the upward trend across all scenarios. Similar to Figure \ref{fig:9}, the gap between TS STAR-RIS and ES STAR-RIS uplink transmission widens as the local processing data size increases, due to the increased challenge of handling inter-interference in ES STAR-RIS scenarios.

\section{Conclusion}
This paper studied the efficient STAR-RIS operation mode in uplink and downlink transmissions for minimizing energy consumption in a WPT-FL multi-antenna AP system with NOMA. To allocate resources such as time slots, power for harvesting, uplink and downlink broadcasting, computation frequency for users, transmission and reflection phase shift vectors of STAR-RIS, and beamforming matrices for all phases, we formulated a non-convex energy minimization problem for each scenario, depending on the STAR-RIS operation mode in uplink and downlink transmissions.

Due to the coupling effect among variables, we decoupled the original optimization problem into two subproblems in each scenario. In the first subproblem, the transmission and reflection phase shift vectors of STAR-RIS, along with the beamforming matrices for energy harvesting, uplink, and downlink transmission phases, were jointly optimized using BCD over SDP or the sum of RQ. Subsequently, two ODSAs in series optimized resources for all phases in ES-ES and TS-ES scenarios, while the ODSA, with a combined BA and ODSA, allocated the resources in ES-TS and TS-TS schemes.

Numerical results demonstrated that the TS-TS scenario achieves the highest energy savings, followed by TS-ES, ES-TS, ES-ES, and the conventional RIS. This superiority of scenarios with TS in uplink transmission was attributed to the absence of inter-user interference and its capability to flexibly adapt STAR-RIS phase shift vectors for each group during uplink and downlink transmission. In contrast, scenarios with ES in the uplink transmission suffered from both intra- and inter-user interference and shared STAR-RIS phase shift vectors for both groups simultaneously, which negatively impacted their performance.

Our paper can be extended in several interesting future directions, such as exploring the integration of RSMA to further mitigate interference, studying robust designs under imperfect CSI, jointly optimizing FL convergence parameters, and leveraging AI-based methods such as reinforcement learning to enable adaptive STAR-RIS mode selection and dynamic resource allocation, thereby supporting more practical and large-scale deployments. Beyond the theoretical design, it is also important to address practical implementation challenges of STAR-RIS in FL systems, including hardware limitations, such as the need for affordable, high-resolution phase shifters, and the complexity of jointly optimizing transmission and reflection coefficients.

\bibliographystyle{IEEEtran}
\bibliography{IEEEabrv,biblio}

\begin{IEEEbiography}[{\includegraphics[width=1in,height=125in, clip,keepaspectratio]{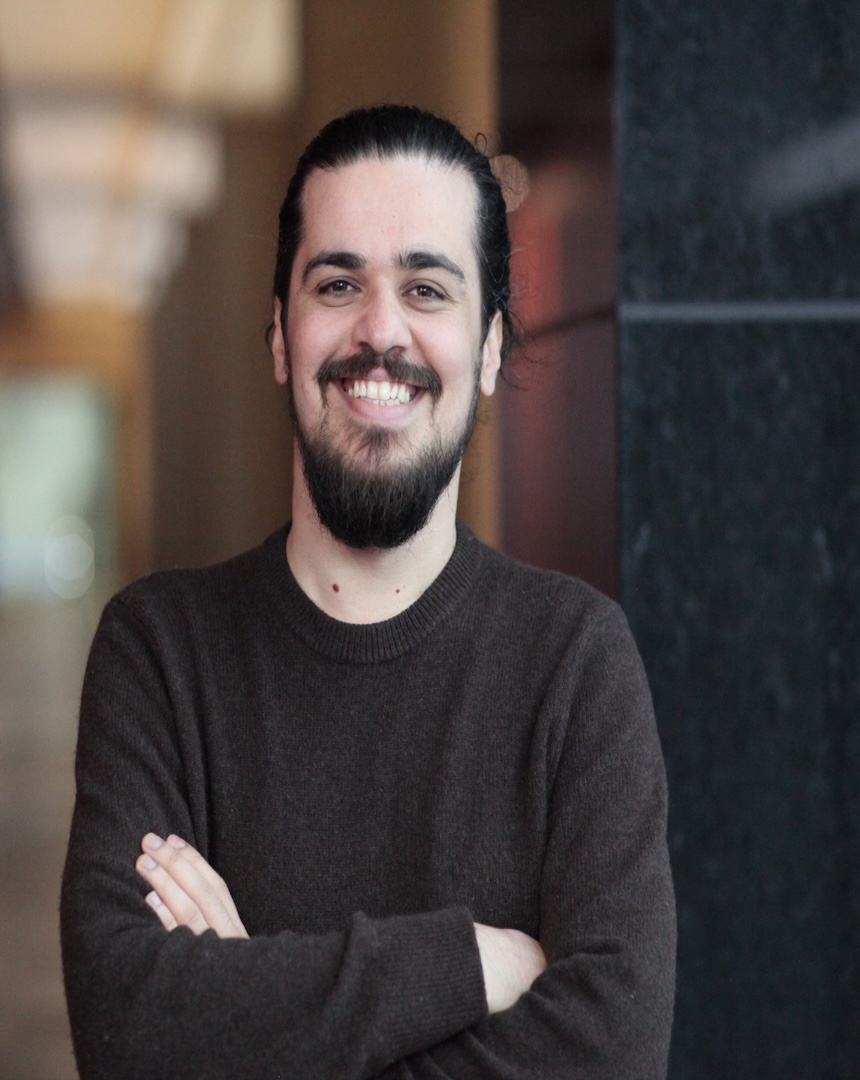}}]{MohammadHossein Alishahi} received the B.E. degree in Communication Systems from the University of Birjand, Iran, and the M.Sc. degree in Communication Systems Engineering from the Iran University of Science and Technology (IUST), Tehran, Iran. He received the Ph.D. (with honors) degree in Communication Systems from Laval University, Québec, Canada, in 2025. His doctoral research focused on the optimization and design of RIS for beyond 5G/6G networks, with contributions to spectral and energy efficiency improvements. He is currently a Postdoctoral Fellow at the University of Toronto, Canada. His research interests include beyond 5G/6G networks, convex and non-convex optimization, interference alignment, and wireless power transfer. He also serves as a reviewer for leading IEEE journals, including IEEE Transactions on Wireless Communications, IEEE Transactions on Communications, and IEEE Transactions on Vehicular Technology.
\end{IEEEbiography}

\begin{IEEEbiography}[{\includegraphics[width=1in,height=1.25in,clip,keepaspectratio]{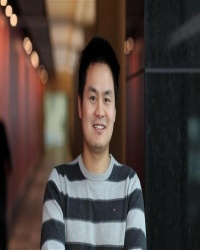}}]{Ming Zeng}
 (ming.zeng@gel.ulaval.ca) [M’ 2019] is currently an Associate Professor and a Canada Research Chair with the Department of Electrical and Computer Engineering, Laval University, Quebec City, QC, Canada. He received the B.E. and master’s degrees from Beijing University of Post and Telecommunications, Beijing, China, in 2013 and 2016, respectively, and the Ph.D. degree in telecommunications engineering from Memorial University of Newfoundland, St John’s, NL, Canada, in 2020. He has published more
than 100 articles and conferences in first-tier IEEE journals and proceedings,
and his work has been cited over 5400 times per Google Scholar. His research
interests include resource allocation for beyond 5G systems and machine
learning-empowered optical communications. Dr. Zeng serves as an Associate
Editor for the IEEE Transactions on Communications, IEEE OJ-COMS, and the IEEE Wireless Communications Letters.
\end{IEEEbiography}

\begin{IEEEbiography}[{\includegraphics[width=1in,height=1.25in, clip,keepaspectratio]{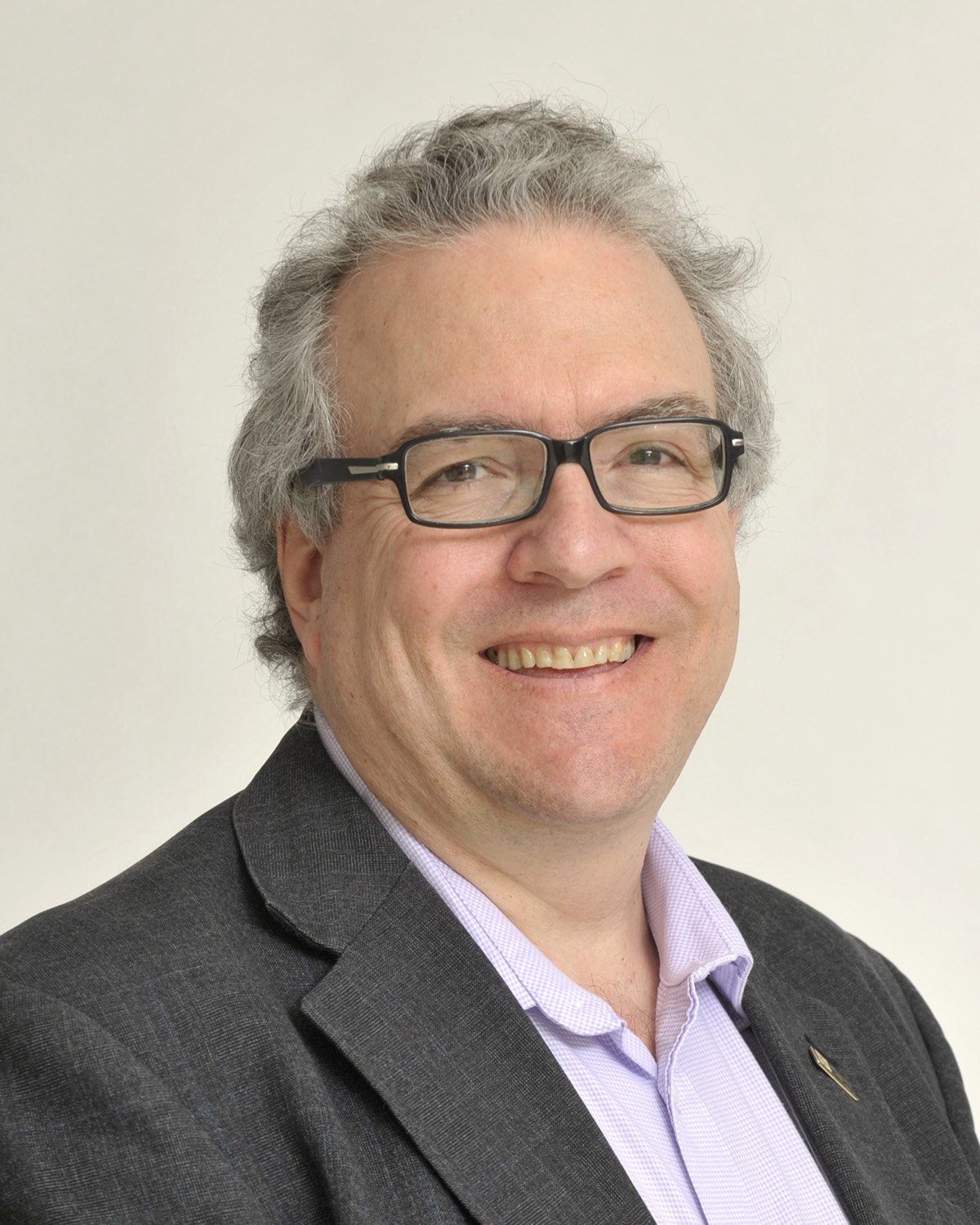}}]{Paul Fortier}  (SM 00) received his B.Sc. and his M.Sc. in electrical engineering from Laval University in 1982 and 1984, respectively and his M.S. in statistics and his Ph.D. in electrical engineering from Stanford University in 1987 and 1989, respectively. Since 1989, he has been with the ECE Department at Laval University, where he is currently a contract adjunct professor. From 1997 to 2003 and from 2016 to 2023, he was Chair of the ECE Department. From 2003 to 2007, he was Associate Dean for Development and Research at the Faculty of Science and Engineering. From 2007 to 2009, he was Vice-President, Scientific Affairs and Partnerships at FRQNT (Quebec’s granting agency). From 2010 to 2012, he was Vice-President for Research and Innovation at Laval University. His research interests include digital signal processing for communication systems and the study of complexity and performance trade-offs in hardware implementations. Paul Fortier is a Fellow of the Engineering Institute of Canada, a Fellow of the Canadian Academy of Engineering, and a Senior Member of the IEEE.
\end{IEEEbiography}

\begin{IEEEbiography}[{\includegraphics[width=1in,height=1.25in, clip,keepaspectratio]{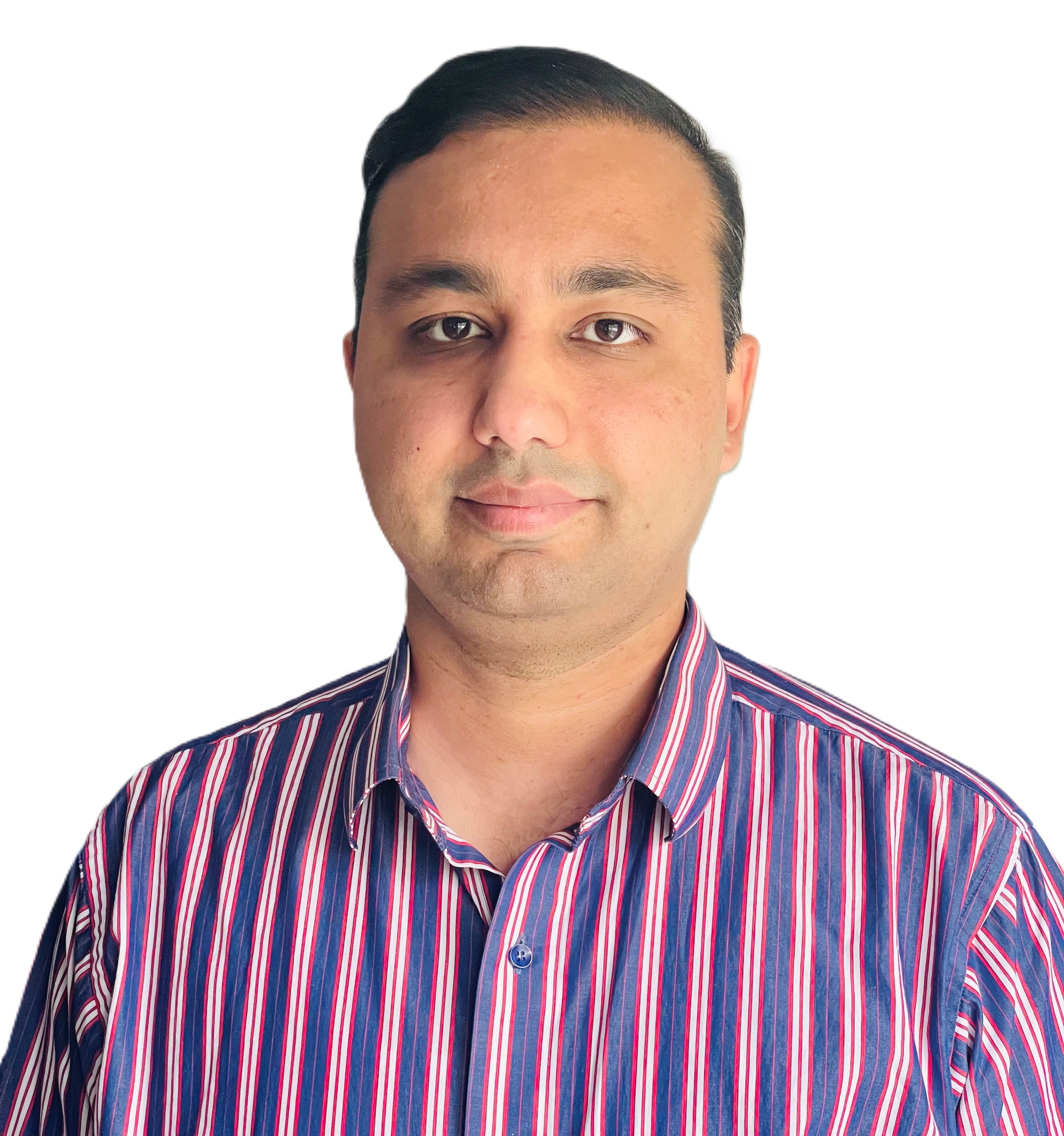}}]{Omer Waqar} (Senior Member, IEEE) received the B.Sc. degree in electrical engineering from the University of Engineering and Technology (UET), Lahore, Pakistan, in 2007, and the Ph.D. degree in electrical and electronic engineering from the University of Leeds, Leeds, U.K., in November 2011. From January 2012 to July 2013, he was a Research Fellow with the Center for Communications Systems Research and 5G Innovation Center, University of Surrey, Guildford, U.K. He worked as an Assistant Professor with UET from August 2013 to June 2018. He worked as a Researcher with the Department of Electrical and Computer Engineering, University of Toronto, Canada, from July 2018 to June 2019. He worked as an Assistant Professor with the Department of Engineering, Thompson Rivers University, Kamloops, BC, Canada, from August 2019 to July 2023. Since August 2023, he has been working as an Assistant Professor with the School of Computing, University of the Fraser Valley, Abbotsford, BC, Canada, and as an Adjunct Faculty with York University, Toronto, ON, Canada. He has authored or coauthored 45+ peer-reviewed articles, including top-tier journals, such as IEEE TRANSACTIONS ON VEHICULAR TECHNOLOGY and IEEE TRANSACTIONS ON MOBILE COMPUTING. His current research interests include reconfigurable intelligent surface-aided communication systems, deep learning for next-generation communication networks, non-convex optimization for 6G networks, and resource allocation of wireless networks for several distributed machine learning paradigms. He has secured over $200K$ in research grants from the Tri-Council Agency, i.e., NSERC Discovery grant and NSERC Alliance grants. He is currently serving as an Associate Editor for the IEEE OPEN JOURNAL OF THE COMMUNICATIONS SOCIETY and IEEE CANADIAN JOURNAL OF ELECTRICAL AND COMPUTER ENGINEERING.
\end{IEEEbiography}

\begin{IEEEbiography}[{\includegraphics[width=1in,height=1.25in, clip,keepaspectratio]{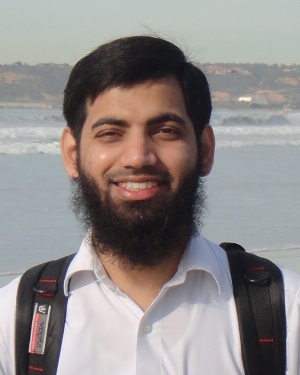}}]{Mohammad Hanif} (Senior Member, IEEE) received the Ph.D. degree in electrical engineering from the University of Victoria, Victoria, BC, Canada, in 2016. He was a postdoctoral fellow at the University of Alberta, Edmonton, AB, Canada, from 2016 to 2018, and at the University of Saskatchewan, Saskatoon, SK, Canada, from 2018 to 2019. He joined Thompson Rivers University in Kamloops, BC, Canada, in 2019 and is currently an Associate Professor in the Department of Engineering at the university. His research interests are in the general area of signal processing and wireless communication systems, including reconfigurable-intelligent-surface-based wireless systems, machine-to-machine communications, molecular communications, and machine learning for future wireless networks.

\end{IEEEbiography}

\begin{IEEEbiography}[{\includegraphics[width=1in,height=1.25in, clip,keepaspectratio]{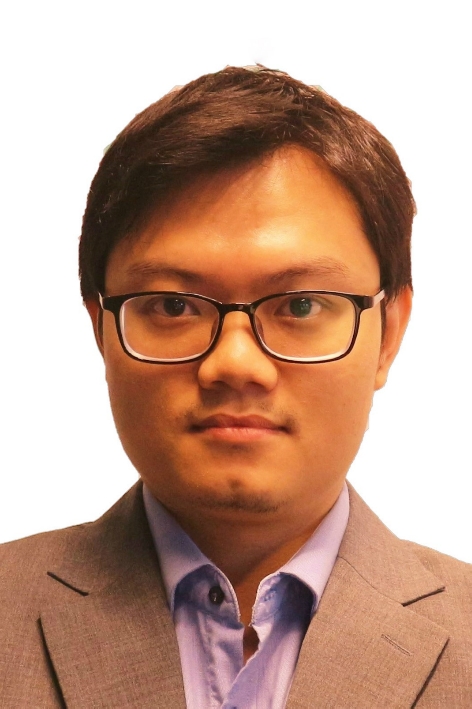}}]{Dinh Thai Hoang} (IEEE Senior Member) (M’16, SM’22) is currently a faculty member at the School of Electrical and Data Engineering, University of Technology Sydney, Australia. He received his Ph.D. in Computer Science and Engineering from the Nanyang Technological University, Singapore 2016. His research interests include emerging wireless communications and networking topics, especially machine learning applications in networking, edge computing, and cybersecurity. He has received several prestigious awards, including the Australian Research Council Discovery Early Career Researcher Award, IEEE TCSC Award for Excellence in Scalable Computing for Contributions on “Intelligent Mobile Edge Computing Systems” (Early Career Researcher), IEEE Asia-Pacific Board (APB) Outstanding Paper Award 2022, and IEEE Communications Society Best Survey Paper Award 2023. He is currently an Editor of IEEE TMC, IEEE TWC, IEEE TCOM, and IEEE TNSE.
\end{IEEEbiography}

\begin{IEEEbiography}[{\includegraphics[width=1in,height=1.25in, clip,keepaspectratio]{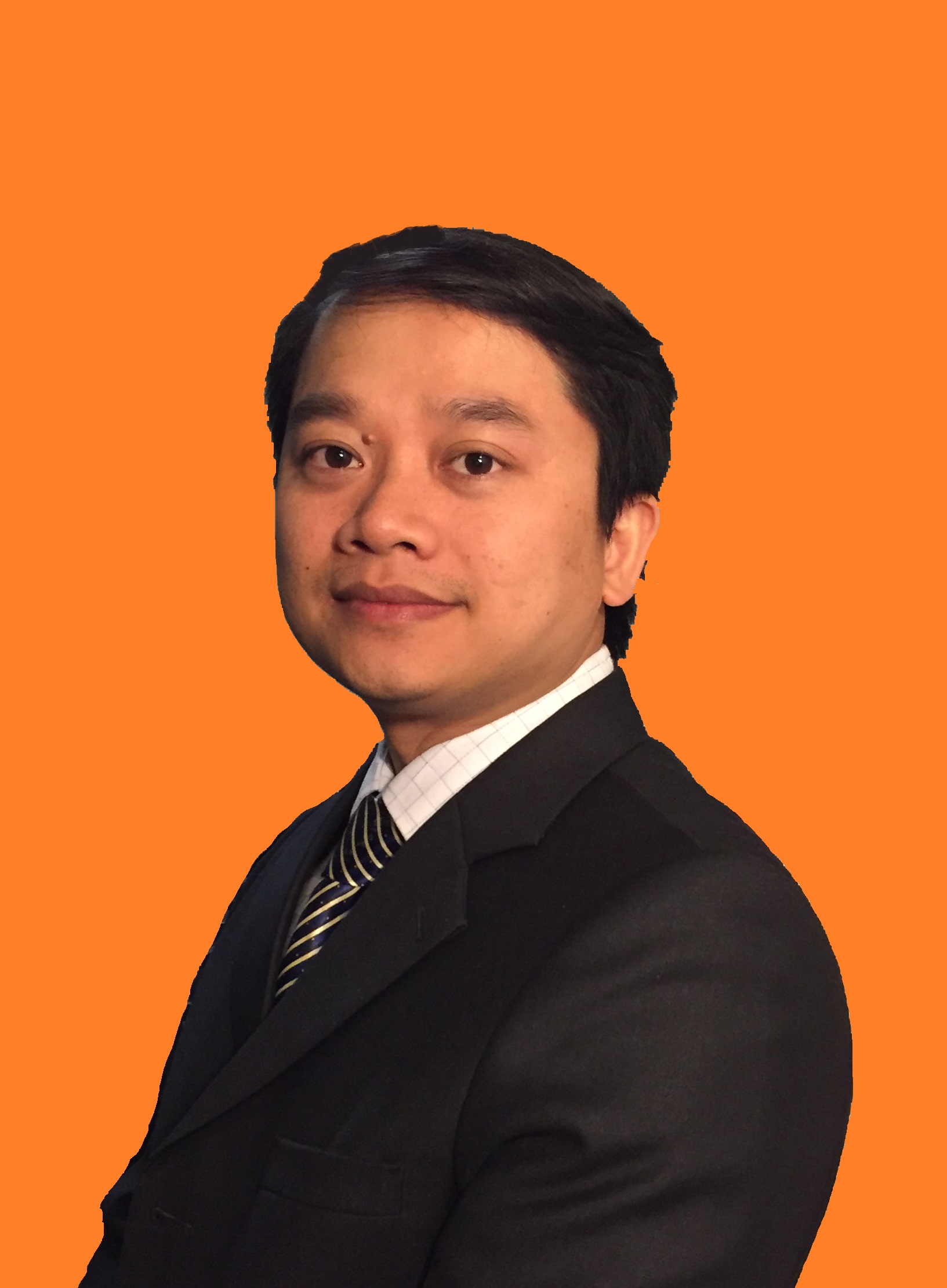}}]{Diep N. Nguyen} (Senior Member, IEEE) received the M.E. degree in electrical and computer engineering from the University of California at San Diego (UCSD), La Jolla, CA, USA, in 2008, and the Ph.D. degree in electrical and computer engineering from The University of Arizona (UA), Tucson, AZ, USA, in 2013. He is currently the Head of 5G/6G Wireless Communications and Networking Lab, Director of Agile Communications and Computing group, Faculty of Engineering and Information Technology, University of Technology Sydney (UTS), Sydney, Australia. Before joining UTS, he was a DECRA Research Fellow with Macquarie University, Macquarie Park, NSW, Australia, and a Member of the Technical Staff with Broadcom Corporation, CA, USA, and ARCON Corporation, Boston, MA, USA, and consulting the Federal Administration of Aviation on turning detection of UAVs and aircraft, and the U.S. Air Force Research Laboratory, USA, on anti-jamming. His research interests include computer networking, wireless communications, and machine learning applications, with emphasis on systems’ performance and security/privacy. Dr. Nguyen received several awards from U.S. Congress, the U.S. National Science Foundation, and the Australian Research Council. He has served on the Editorial Boards of the IEEE Transactions on Mobile Computing, IEEE Communications Surveys \& Tutorials (COMST), IEEE Open Journal of the Communications Society.
\end{IEEEbiography}

\begin{IEEEbiography}[{\includegraphics[width=1in,height=1.25in,clip,keepaspectratio]{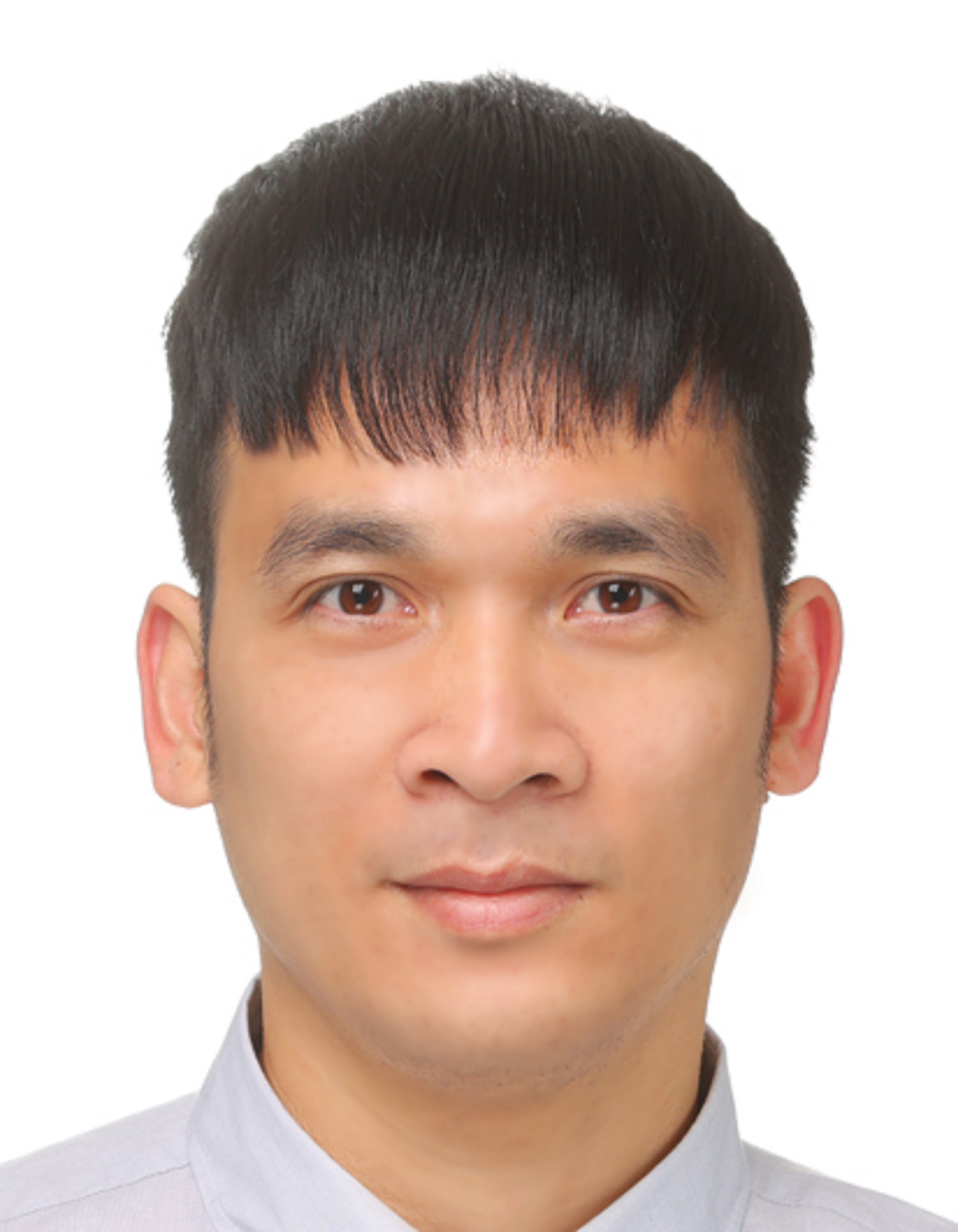}}]{Quoc-Viet Pham} (M’18, SM’23) is currently an Assistant Professor in Networks and Distributed Systems at the School of Computer Science and Statistics, Trinity College Dublin, Ireland. He earned his BSc and PhD degrees (with Best PhD Dissertation Award) in Telecommunications Engineering from Hanoi University of Science and Technology and Inje University in 2013 and 2017, respectively. He has special research interests in the areas of wireless AI, edge computing, Internet of Things, and distributed learning. He was a recipient of the IEEE TVT Top Reviewer Award in 2020, Golden Globe Award in Science and Technology for Vietnam’s Young Researchers in 2021, IEEE ATC Best Paper Award in 2022, and IEEE MCE Best Paper Award in 2023, IEEE M-COMSTD Exemplary Editor Award in 2024, and Clarivate Highly Cited Researcher Award in 2024. He was honoured with the IEEE ComSoc Best Young Researcher Award for EMEA 2023 in recognition of his research activities for the benefit of the Society. He was the Lead Guest Editor of the IEEE Internet of Things Journal special issue on Aerial Computing for the Internet of Things. He currently serves as an Editor of IEEE Communications Letters, IEEE Communications Standards Magazine, IEEE Communications Surveys \& Tutorials, Journal of Network and Computer Applications, IEEE Transactions on Mobile Computing, and IEEE Transactions on Network Science and Engineering.

\end{IEEEbiography}

\end{document}